\documentclass{mn2e}
\usepackage{times}
\input{psfig.sty}
\newif\ifAMStwofonts

\title[X-ray spectra of inhomogeneous accretion flows]
         {On the X-ray spectra of luminous, inhomogeneous accretion flows}
\author[A. Merloni et al.]
         {A. Merloni$^{1}$\thanks{E-mail: am@mpa-garching.mpg.de}, J. Malzac$^{2}$, A.C. Fabian$^{3}$ and
         R.R. Ross$^{4}$\\$^{1}$Max-Planck-Institut f\"ur Astrophysik,
Karl-Schwarzschild-Strasse 1, D-85741, Garching,
Germany\\$^{2}$Centre d'Estude Spatiale des Rayonnements, CNRS-UPS 9,
         Avenue du Colonel Roche, 31028 Toulouse, Cedex 4, France\\$^{3}$Institute of Astronomy, Madingley Road,
         Canbridge, CB3 0HA\\$^{4}$Physics Department, College of the Holy
         Cross, Worcester, MA 01610, USA}

\def\ep{\varepsilon}

\def\tT{\tau_{\rm T}}

\def\tB{\tau_{\rm B}}

\def\pescr{\dot{P_{\rm R}}}
\def\pescs{\dot{P_{\rm UV}}}

\def\pescc{\dot{P_{\rm C}}}

\def\Lh{L_{\rm h}}

\def\Lx{L_{\rm X}}
\def\LR{L_{R}}

\def\Uuv{U_{\rm UV}}

\def\Uc{U_{\rm C}}
\def\UR{U_{R}}
\def\aR{a_{R}}
\begin{document}

\maketitle
\label{firstpage}

\begin{abstract}
We discuss the expected X-ray spectral and variability properties of
black hole accretion discs at high luminosity, under the hypothesis
that radiation pressure dominated discs are subject to violent
clumping instabilities and, as a result, have a highly inhomogeneous
two-phase structure. After deriving the full
accretion disc solutions explicitly in terms of the parameters 
of the model, we study their radiative properties both with a simple
two-zones model, treatable analytically, and with radiative transfer
simulations which account simultaneously for energy balance
 and Comptonisation in the hot phase, together with reflection, reprocessing,
ionization and thermal balance in the cold phase.
We show that, if not only the density, but also the heating
rate within these flows is inhomogeneous, then complex reflection-dominated
spectra can be obtained for a high enough covering fraction of the cold
phase. In general, large reflection components in the observed X-ray
spectra should be associated with strong soft excesses, resulting
from the combined emission of ionized atomic emission lines. 
The variability
properties of such systems are such that, even when contributing to a
large fraction of the hard X-ray spectrum, the reflection component is
less variable than the power-law like emission originating from the
hot Comptonising phase, in agreement with what is observed in many Narrow
Line Seyfert 1 galaxies and bright Seyfert 1. Our model falls within
the family of those trying to explain the complex X-ray spectra of
bright AGN with ionized reflection, but presents an alternative,
specific, physically motivated, geometrical setup for the complex multi-phase
structure of the inner regions of near-Eddington accretion flows.
\end{abstract}


\section{Introduction}

Black Hole systems, either in X-ray
binaries (BHXRB) or in active galactic nuclei (AGN),
when shining at luminosities close to
the Eddington limit are thought to be powered by accretion
through geometrically thin, optically thick discs, where angular
momentum is transferred by the stresses due to magneto-rotational
instability (MRI; Balbus \& Hawley 1991, 1998).
 From the observational point of view, the presence of such discs
is inferred from the spectral properties of
transient BHXRB at (or close to) the peak of their
outburst luminosities (in the so-called High/Very High/Intermediate
states, 
see e.g. Belloni 2004, or McClintock \& Remillard 2005 for a different
nomenclature, and
references therein) and from the dominance of the quasi-thermal UV
emission in bright Quasars (see e.g. Malkan
1983; Laor 1990; Shang et al. 2005). According to the standard models
(Shakura \& Sunyaev, 1973), such accretion discs are radiatively
efficient, as they are able to convert into radiation a sizeable
fraction (between 6 and $\sim$40\%, depending on black hole spin)
of the available gravitational
energy of the accreting gas.
Indeed, radiative efficiency
estimates obtained by comparing the
total supermassive black hole mass density in the local
universe with the accretion
power released by Quasars and AGN suggest that this is
probably the main mode of growth by accretion of supermassive black
holes (see e.g. Fabian \& Iwasawa 1999; Elvis, Risaliti and Zamorani
2002; Yu and Tremaine 2002; Marconi et al. 2004; Merloni, Rudnick \&
Di Matteo 2004). It is
therefore clear that a good understanding of the accretion mechanism
associated with nearly-Eddington sources can shed light on fundamental
issues of black holes astrophysics.

Nonetheless, according to the above mentioned
  standard accretion disc solutions, such
  highly luminous discs should be
  radiation pressure dominated and therefore
  unstable to perturbations of both mass flow (Lightman \& Eardley
  1974) and heating rates (Shakura \& Sunyaev 1976).
Thus, it is not clear yet to what extent this standard solutions
  represent a realistic description of the observed systems.
In recent years, both analytic works
(Blaes and Socrates 2001, 2003)
and simulations (Turner, Stone and Sano 2002, Turner et al. 2003b,
Turner 2004) have shown that magnetized,
radiation pressure dominated accretion discs may be in fact
subject to violent clumping instabilities if magnetic pressure exceeds
  gas pressure and photons are able to diffuse from compressed regions.
Large density variations may also be caused by photon bubble instabilities
(Gammie 1998), which may develop into a series of shocks propagating
  through the plasma (Begelman 2001,2002; Turner at al. 2005).
These instabilities may in turn have profound
effects not only on the nature of the cooling mechanism of luminous
  discs, but also on their observational appearance (see e.g., Davis
  et al. 2004; Ballantyne  et al. 2004), as we discuss in this work.

Radiative models for inhomogeneous, clumpy discs that explain the X-ray
properties of accreting black holes have been
proposed since long time. Most of the previous works were devoted to
the study of the internal state  of the cold clumps and on their
radiative output (Guilbert \& Rees 1988; Celotti, Fabian \&
Rees 1992; Collin-Souffrin et al. 1996; Kuncic, Celotti \& Rees 1997).
Krolik (1998) studied analytically the
overall equilibrium of a clumpy two-phase accretion disc, while Malzac
\& Celotti (2002; MC02) studied numerically the Comptonization equilibrium
between the cold clumps and the hot surrounding plasma and calculated
the emitted X-ray spectrum.

In this paper, we explore in detail the consequences of the hypothesis
that accretion flows close to the Eddington rate are indeed 
inhomogeneous.
In particular, we present a detailed study of their expected X-ray
spectral and variability properties, extending
the results of MC02 both numerically and
analytically. From the numerical point of view, we calculate the 
emerging
spectrum from an inhomogeneous, two-phase, clumpy flow by coupling the
non-linear Monte Carlo code of Malzac \& Jourdain (2000), that
calculates the Compton equilibrium between the cold and the hot phase, with the X-ray
ionized reflection code of Ross \& Fabian (1993, 2005), that accurately
computes the reprocessed radiation in the cold phase.
 From the analytic point of view, we are able to show that a crucial 
factor in
determining the broad band spectral properties of an inhomogeneous flow
is not only the amount of cold clouds/filaments pervading the hot
plasma (as considered in all the previous studies cited above), but
also the level of inhomogeneity of the {\it heating} in the hot phase
itself. Indeed, we will demonstrate both analytically and numerically,
the very important result that the more localized the heating is
(as due to, e.g. a reconnection event), the more the emerging spectra
are dominated by reflection/reprocessing, as already suggested by
Fabian et al. (2002).

In this paper we will also show that a model for 
an accretion flow
which is inhomogeneous both in its density and heating structure
can explain many of the observed properties of black
holes accreting at rates
close to the Eddington one (e.g. bright QSOs and Narrow Line Seyfert galaxies). 
In particular, flows with
varying degrees of inhomogeneity
naturally produce little variability in the observed reflected
component associated with large variations of the power-law continuum,
and reflection dominated spectra at the lowest-luminosity levels, in a
strikingly similar fashion to what predicted by the so-called
light-bending model (Miniutti \& Fabian 2004).
Similarly, such a configuration naturally produces multiple
reflections of the X-ray emission that can enhance the  broad
emission and absorption features in the observed spectra
(Collin-Souffrin et al. 1996; Ross, Fabian
\& Ballantyne, 2002).

The paper is structured as follows: In Section~\ref{sec:toy} we
first introduce our simple toy model for a clumpy, two-phase medium
with inhomogeneous heating. As this model is treatable analytically, we
are able to discuss the main expected properties (spectra and
variability) over a wide area  of the parameter space. In
Section~\ref{sec:simu} we then discuss our numerical simulations,
while in Section~\ref{sec:discus} we discuss the relevance of
our results for observations of high accretion rate BHXRB and AGN and
their implications for models of accretion flows. Finally, we
draw our conclusions in Section~\ref{sec:conc}. In addition, we have
developed in Appendix~\ref{app:model} a full inhomogeneous accretion
disc model based on the picture discussed in Section~\ref{sec:toy}
in which the dependence of our model parameters on radius,
black hole mass and accretion rate is made explicit.

\section{A simple toy model}
\label{sec:toy}
In order to gain insight into the main expected observable properties of
inhomogeneous discs, we introduce first a simple, yet motivated,
idealized model,
which is tractable analytically. In Appendix~\ref{app:model},
instead, we present a full inhomogeneous accretion  disc model,
closely following the ideas of Krolik (1998), with the aim of
constraining the values and the scalings of the parameters we
introduce in order to describe the disc inhomogeneity (see below).

Let us consider a sphere of hot plasma with uniform density and radius
$H$, pervaded by small lumps  of cold matter randomly distributed
throughout the hot phase. Such a system may
represent the inner region of an accretion flow (with $H \approx$ few
Schwarzschild radii, $R_{\rm S}$),
or a patch of the accretion disc itself
in the radiation pressure dominated region, where clumping instability
is at work (with $H \approx$ disc scaleheight, see
Appendix~\ref{app:model}). 
Then we can write
$H=rR_{\rm S}=3 \times 10^{12} m_{7} r_{0}$ cm, where $m_{7}$ is the mass
of the central black hole in units of $10^7 M_{\odot}$, and
$r_{0}=r/1$.\footnote{Throughout the paper we use the representation
$A_x=A/10^{x}$.}
As already discussed in MC02, the lack of strong absorption features
in the X-ray spectra of accreting black holes suggests that any
putative cold clump of matter immersed in the hot plasma should be
optically (Thomson) thick (see also, Kuncic et al. 1997),
with zero transmission.
Thus we consider clouds with column densities $\ga 10^{25}$ cm$^{-2}$,
which implies a density $n_{\rm cl}\ga 3.3 \times 10^{13}
(m_{7} \epsilon_{-1} r_0)^{-1}$ cm$^{-3}$, where we have introduced
the parameter $\epsilon=0.1 \epsilon_{-1}$, the ratio of the cloud
size to the system size $H$.

\begin{figure}
\psfig{figure=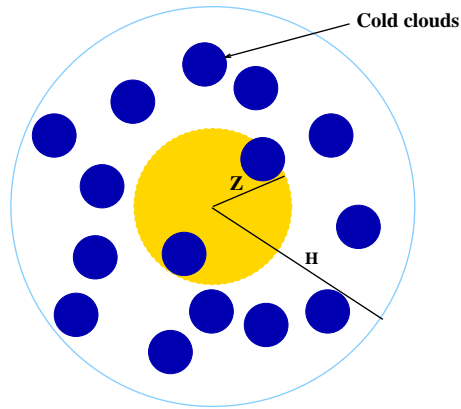,angle=270,width=0.5\textwidth}
\caption{A simple two-zone model for an inhomogeneous accretion 
 flow. The hot plasma emitting the hard X-rays is pervaded by small
clouds of dense matter, optically thick to Thomson scattering. Both
cold clumps and the hot plasma are uniformly distributed, but the heating is
   only concentrated inside the inner sphere of radius $Z\equiv h H$. The system 
is assumed to be in radiative equilibrium.}
\label{fig:setup}
\end{figure}

We assume that the whole system is in radiative equilibrium. 
However, differently from MC02, we allow the heating to be
localized in a small region, rather than diffused throughout the hot
phase. For simplicity, we assume that the heating of the hot plasma
takes place only in an inner concentric sphere of radius $Z\equiv h H$
(see Fig.~\ref{fig:setup}), where a power $L_{\rm h}$ is deposited.
The outer zone can therefore be viewed as a passive reprocessor
(see Appendix~\ref{app:toy}).
In our idealized model the injected  power eventually escapes the
system as three different kinds of radiation:
\begin{equation}
L_{\rm h}=L_{\rm UV}+L_{\rm C}+L_{\rm R};
\end{equation}
respectively the UV/soft X-ray reprocessed luminosity, the
Comptonized (power-law like) luminosity and the reflected luminosity.
In our calculation we also assume that the UV/Soft X-ray radiation
reprocessed by the clouds is the only source of soft photons (no
internal dissipation in the cold clumps, see e.g the discussion in
Collin-Souffrin et al. 1996, or Malzac 2001).

\begin{figure}
\psfig{figure=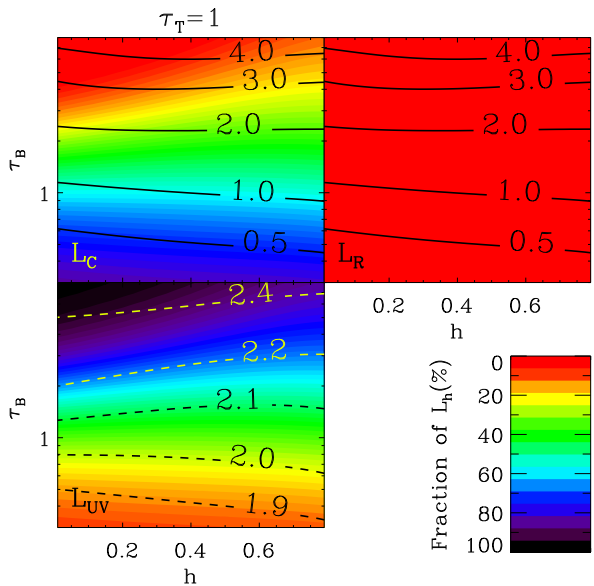,width=0.5\textwidth}
\caption{The relative intensity of the emerging hard X-ray Comptonized 
($L_{\rm C}$, upper
   left panel),
   reprocessed UV ($L_{\rm UV}$, lower left panel), and reflected   
($L_{\rm R}$, upper right panel)
luminosities is plotted as color-coded contours in the
   $(\tau_{\rm B},h)$ plane. Superimposed on them, in the two upper 
panels are the contours of
constant reflection fraction (solid lines), ranging from $R=0.5$ to 
$R=4$,
while in the lower panel we plot the contours (dashed lines)
of constant power-law indexes $\Gamma$, from 1.9 to 2.4. The optical
depth of the hot phase is assumed to be unity $\tau_{\rm T}=1$}
\label{fig:lx}
\end{figure}

If the clouds are sufficiently small, we can use an effective cloud
optical depth, $\tau_{\rm B}$, as the main parameter controlling the
amount of cold matter. This is defined in such a way that
a photon crossing the hot
plasma has a probability $1-e^{-\tau_{\rm B}}$ of intercepting a 
cloud.
The second parameter that affects the emerging spectrum is the optical
depth to Thomson scattering of the hot plasma, $\tau_{\rm T}$.
Comptonization models for the X-ray emission from compact objects
typically constrain the hot plasma Thomson optical depth $\tau_{\rm
   T}\sim 1$ (see e.g. Zdziarski 1999). This is also consistent with the
upper and lower limits on $\tau_{\rm T}$ derived by Krolik (1998) from the
thermal and evaporative equilibrium of a clumpy disc.

A rough order of magnitude estimate of the expected 
ionization parameter can be derived
as follows: radiatively heated clouds with no internal dissipation,
if dense enough (as we are assuming here, see also Collin-Souffrin et
al. 1996; Kuncic et al. 1997), will re-emit the absorbed flux as a
black-body with typical energy $k T_{\rm bb} \simeq 30 L_{44}^{1/4}
(m_7 r_0)^{-1/2}$ eV, where $L$ is the bolometric luminosity of the
system and the weak dependence on the albedo has been neglected. 
Then, assuming that in the hot phase gas pressure dominates over
radiation pressure (an assumption that can be verified a posteriori,
see details in Appendix~\ref{app:model}), 
thermal pressure balance between the cold clouds and the
hot phase implies that
\begin{equation}
n_{\rm cl} \simeq 3.3 \times 10^{15} \left(\frac{\tau_{\rm T}}{1}\right)
\left(\frac{kT_{\rm hot}}{100 \;{\rm keV}}\right) (m_7 r_0)^{-1/2}
L_{44}^{-1/4} \; {\rm cm}^{-3}
\end{equation}
Therefore, the typical ionization parameter\footnote{Here, as in the
  remaining of the paper, the ionization parameter will be expressed in
  erg cm s$^{-1}$.}
\begin{equation}
\xi=4\pi F/n_{\rm cl}\simeq 3.5 \times 10^3 L_{44}^{5/4} (m_7
r_0)^{-3/2}  \left(\frac{\tau_{\rm T}}{1}\right)^{-1} \left(\frac{k
T_{\rm hot}}{100\;{\rm keV}}\right)^{-1}.
\end{equation}
It is worth stressing, however,  that in the inhomogeneous heating
model we are discussing, a range of ionization parameters is expected,
given the sensitivity of $\xi$ to the actual geometry of the
system. The above equation has to be considered just an order of
magnitude estimate. Indeed, the above estimates are consistent with
the values obtained with the full disc modeling outlined in
Appendix~\ref{app:model}, where the explicit scalings of cloud
density, temperature and ionization parameters with accretion disc
radius, black mass and accretion rate are presented.
In the following (see Section~\ref{sec:simu}), we will explore
the effects of changing the value of $\xi$ on the spectra emerging
from our inhomogeneous flow.

The other parameters needed to compute the emerging
radiation intensity are the energy and angle-integrated
single-reflection albedos for a
Comptonized and reflection spectrum, $a$ and $a_{\rm R}$,
respectively. Malzac, Dumont \& Mouchet (2005) have shown that,
when the reprocessed spectrum is computed self-consistently
the X-ray albedo is a strong function of the ionization parameter and of the
spectrum of the illuminating radiation. Moreover, multiple reflections
start to become important for $\tau_{\rm B}\sim 1$. In a simple
one-zone model, the net albedo produced by multiple reflections should
be approximately given by (Eq. 2 of Ross, Fabian and Ballantyne 2002): 
\begin{equation}
a_{\rm net}=\ep a/[1-(1-\ep)a],
\label{eq:anet}
\end{equation} 
where $\ep$ is the escape probability for a reflected photon and $a$
is the albedo for a single reflection by the material. 
We should expect that, whenever heating is inhomogeneous 
the cold clumps in the inner heated zone and in
the outer one will not only be exposed to illuminating radiation with
different spectra, but also experience different numbers of multiple
reflections. Thus, it is plausible to conceive that in general $\ep$ will not
be identical to
$\exp(-\tau_{\rm B})$ as would be in a simple one zone model.
For the sake of simplicity, and because the main purpose
of the toy model discussed here is to give a coherent qualitative
picture of the properties of inhomogeneous flows, we have chosen 
to calibrate the net albedo directly on the
simulated spectra that will be presented in Section~\ref{sec:simu}.
To a very high degree of accuracy, we find that, for a ionization
parameter of $\xi=300$, the net albedo is a function of the
optical depth of the cold clumps only, and can be expressed with a similar
expression by replacing $\ep=\exp(-\tB/1.43)$, with $a=0.36$ (see
also figure~\ref{fig:albedo}). 
The reflected radiation can itself be reflected many times for the
most extremely inhomogeneous discs. Also for its albedo we have used
the expression (\ref{eq:anet}) with $\ep=\exp(-\tB/1.82)$, with
$a_{\rm R}=0.4$. This latter value is also consistent with the Monte Carlo
simulations of Malzac (2001), where the albedo for a typical
reflection spectrum illuminating neutral matter was computed. We have
also verified that assuming an increased albedo of $a_{\rm R}=0.6$ to
account for the effects of ionization have negligible effects on the
final results of our analytic two-zones model.

We can then proceed as in Section 4 of MC02 and solve for the three 
different luminosities
$L_{\rm UV}$, $L_{\rm C}$ and $L_{\rm R}$ (and the corresponding
reflection fraction $R$) making use of the analytic
approximations for the corresponding photon escape  probability
calculated from Monte Carlo simulations (see Eqs. A3 and A5 in MC02).
We keep the heating rate, $L_{\rm h}$, fixed and
explore the parameter space spanned by $0.1<\tau_{\rm B}<8$ and $0.05<h<0.9$.
Details of the analytic calculation of the emerging
luminosities in this two-zone model are given in the Appendix~\ref{app:toy}.
Once the soft and the Comptonized luminosities are found, we determine the
spectral shape of the X-ray emission by calibrating the power-law index
$\Gamma$ with the numerical simulations (see below), where the power
law index of the Comptonized emission is calculated by fitting the
emergent spectra with the {\tt PEXRAV} 
model in the 2-30 keV range. Apart from a very
weak dependence on the heating inhomogeneity parameter, $h$, 
the power-law index depends only logarithmically on $\tB$, 
with the spectrum being softer for larger
cold clouds column densities. From the analysis of the results of the
simulations, we derive the following
phenomenological expression for the power-law index as a function of
$\tau_{\rm B}$: $\Gamma \simeq 2 + 0.4 \log \tB$. 

Figure~\ref{fig:lx} shows the emerging luminosities as coloured contours in the
two-dimensional parameter space $(\tau_{\rm B},h)$, for a 
Thomson optical depth of the hot phase $\tau_{\rm T}=1$. Shown are the
relative intensities (normalized to the total heating rate $L_{\rm h}$) 
of the X-ray Comptonized luminosity ($L_{\rm C}$, upper left panels), of
the reflected luminosity ($L_{\rm R}$, upper right panels) and of the
reprocessed UV/soft X-ray one ($L_{\rm UV}$, lower left panels).
Superimposed on them, in the two upper panels are the contours of
constant reflection fraction, defined as $R=L_{\rm R}/(a_{\rm net} L_{\rm
  X})$, where $L_{\rm X} \equiv L_{\rm C}(E>1{\rm keV})$ is the Comptonized
luminosity emerging above 1 keV. This has been calculated analytically
from $L_{\rm C}$ assuming that the Comptonized emission is power-law
with index $\Gamma$ extending from $kT_{\rm bb}=50$ eV up to an energy
corresponding to the temperature of the hot electrons $\Theta\equiv kT_{\rm
  hot}/(m_{\rm e}c^2)\approx (3/19)/[(\Gamma-1)\tau_{\rm T}^{4/5}] $ 
(Wardzi{\'n}ski and Zdziarski 2000).
In the lower panel we plot the contours of constant power-law
spectral index, $\Gamma$.

As expected,
the reflection fraction $R$ increases with  $\tau_{\rm B}$, as does
the relative intensity of the soft reprocessed emission: the larger
the filling factor of the cold component, the more reflection
dominated the emerging spectrum will be. In general, large reflection
fractions are always expected when the reprocessed component is strong
compared to the direct Comptonized X-ray emission.

The relative intensity of the soft reprocessed luminosity, $L_{\rm UV}$,
is larger in the
centrally heated case than in the uniformly heated case, 
if $\tau_{\rm B}$ is large.
As a result, for large enough
$\tau_{\rm B}$ and localized heating (i.e. in the upper left corner
of the plots shown in Fig.~\ref{fig:lx}),
the X-ray continuum power-law is dominated
by the reflection component and by the thermal reprocessed emission.

\begin{figure}
\psfig{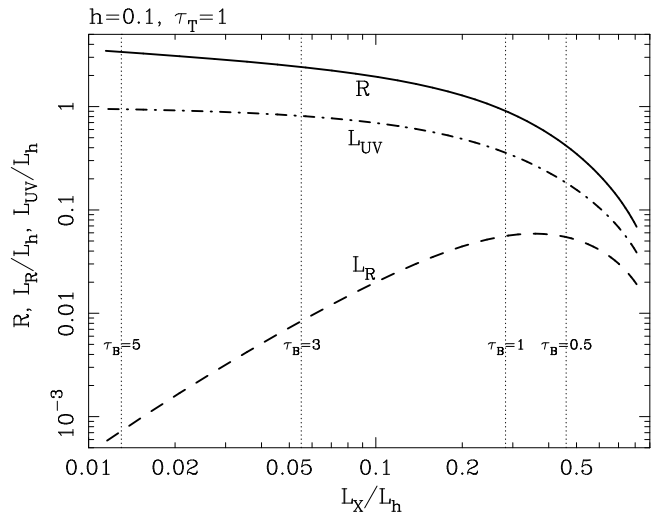}
\caption{The relative intensity of the reflection component ($R$,
   solid line), of the
   reprocessed UV ($L_{\rm UV}$, dot-dashed line) and  reflected
   ($L_{\rm R}$, dashed line) luminosities are plotted as
  functions of the emergent X-ray (Comptonised power-law)
 luminosity above 1 keV ($L_{\rm X}$) for a varying
   cloud optical depth $\tB$.
   All luminosities are renormalized by the total heating rate $L_h$, which
   is here assumed to be concentrated in an inner sphere of radius
   $h=0.1$. The vertical dotted lines mark the X-ray luminosities
   corresponding to $\tB$=0.5,1,3,5.}
\label{fig:h01}
\end{figure}

\subsection{On the variability of the reflection component}
\label{sec:varref}
One more interesting fact emerges from the study of this simple
two-zone toy model. As can be seen in Fig.~\ref{fig:lx}, the reflected
luminosity $L_{\rm R}$ does not vary as much as $L_{\rm X}$ and
$L_{\rm UV}$ across the explored region of the parameter space, at
least as long as $\tB$ is smaller than $\sim$ 2 (note
that the color scale of Fig.~\ref{fig:lx} is linear).
In fact, the reflected luminosity $L_{\rm R}$ has a broad
maximum almost coincident with the locus of $R \approx 1$.
To make this clearer, we have plotted in Fig.~\ref{fig:h01} the reflected luminosity,
the soft reprocessed luminosity and the reflection fraction as a
function of the X-ray (Comptonized) luminosity above 1 keV for the
case of inhomogeneous heating $h=0.1$ and $\tau_{\rm
   T}=1$ (those are  also the values of the parameters adopted in our
numerical simulations of Section~\ref{sec:simu} below).
The fact that the reflected luminosity has a maximum, implies that
large variations of the emergent X-ray luminosity, $L_{\rm X}$,
associated with changes in the cold clump integrated optical depth
correspond to only modest variations of the reflection component, at least as
long as $L_{\rm X}/L_{\rm h} \ga 0.05$.
On the other hand, for low values of the Comptonized X-ray luminosity,
the reflected
luminosity correlates with $L_{\rm X}$,
  while at high $L_{\rm X}$, the trend is the opposite.

This global behaviour of $L_{\rm R}$ vs. $L_{\rm X}$ 
is analogous to that of the
so-called ``light bending'' model (Miniutti \& Fabian 2004), which has
been previously invoked to explain the variability properties of the
continuum and of the iron line in a
number of Narrow Line Seyfert 1 (NLS1) galaxies and galactic black 
holes in
the very high state (see, e.g. Miniutti et al. 2003; Miniutti, Fabian
\& Miller 2004; Fabian et al. 2004; Rossi et al. 2005). In particular,
the defining property of the reflected luminosity to show a broad
maximum as a function of the emerging X-ray (Comptonized) one is a
common characteristic of both models. We will return to this
similarity later on.

We have explored the variability property of our simple toy model in
greater detail. In order to do so, we have computed the
fractional variability $\sigma_{i}$ of each spectral component
($i={\rm X,UV,R}$),
associated with variations $\delta \log \tau_{\rm B}, \delta \log h$
of the cold clouds optical depth and of the inner heated region size,
respectively:
\begin{equation}
\sigma_{i}(\delta \log \tau_{\rm B}, \delta \log h)\equiv
|\frac{\partial \log L_i}{\partial \log \tau_{\rm B}}|\delta
\log \tau_{\rm B}+|\frac{\partial \log L_i}{\partial \log h}|
\delta \log h \,. \nonumber\\
\label{sig}
\end{equation}

\begin{figure}
\psfig{figure=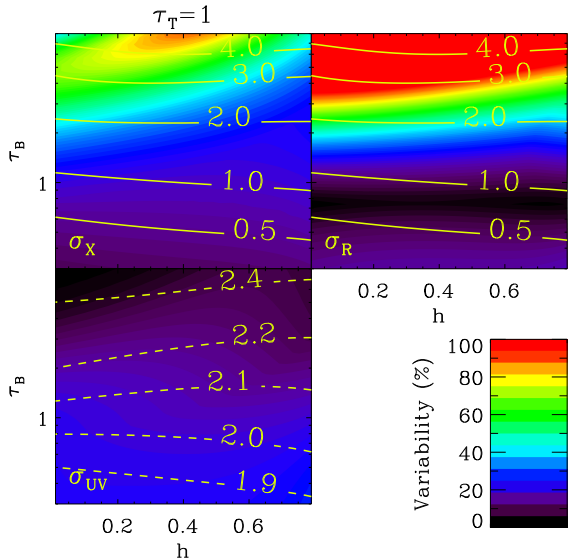,width=0.48\textwidth}
\caption{The total  variability fraction, associated with variations
   of 10\% of $\tau_{\rm B}$ and $h=Z/H$, of Comptonized X-rays emission
   above 1 keV ($\sigma_{\rm X}$, upper
   left panel),
   reprocessed UV ($\sigma_{\rm UV}$, lower left panel), and reflected
   ($\sigma_{\rm R}$, upper right panel)
   luminosities is plotted as color-coded contours in the
   $(\tau_{\rm B},h)$ plane. Superimposed on them, in the two upper
panels are the contours of
constant reflection fraction (solid lines), ranging from $R=0.5$ to
$R=4$,
while in the lower panel we plot the contours (dashed lines)
of constant power-law indexes $\Gamma$, from 1.9 to 2.4, which are the
   same as in Fig.~\ref{fig:lx}.}
\label{fig:var}
\end{figure}

Figure~\ref{fig:var} shows as contour plots the fractional variability
levels of the three components calculated for $\delta \log \tau_{\rm
   B}=\delta \log h= 10\%$.
The fractional variability level of the X-ray emission
can thus be relatively high (up to $\sim$ 50\% for $\tau_{\rm B} \sim 2$)
even for relatively small variations of the total optical depth. 
According to our simple toy model of spherical clouds
in a spherical volume, the total number of clouds, $N$, is related to 
the
cold phase optical depth by $\tau_{\rm B}=(3/4)N\epsilon^2$. The total
number of clouds will have a dispersion around its mean value of $\sim
\sqrt{N}$, so that $\Delta \tau_{\rm B}/\tau_{\rm B} =0.1$ corresponds
to $N\sim 100$ and $\epsilon \sim 0.11 \sqrt{\tau_{\rm
     B}}$. Interestingly, one obtains a similar constrain on the
cloud size from considering thermal equilibrium in the presence of
heat conduction (see Krolik 1998, eq. 5): $\epsilon \simeq 0.2
\tau_{\rm T}^{-1} (kT_{\rm hot}/100\;{\rm keV})^{3/2}$.

Overall, the fractional variability of $L_{\rm X}$ is
higher if the reflection fraction $R$ is high. For $\tau_{\rm B}\la 2$,
and $R<2$, the variability of the Comptonised emission, $\sigma_{\rm 
X}$,
is more than twice that of the UV/soft X-ray and reflected components.
For even lower values of the cold phase optical depth $\tau_{\rm B}\la
0.5$, and
correspondingly lower reflection fractions, the amount of variability
is comparable in both hard and soft, reprocessed,
components, while for the opposite extreme case of very large $\tB$ (i.e. for
the case in which most of the X-ray and reflected emission is hidden from the
view), the reflection component at high energies will vary most dramatically,
while the UV/soft X-ray emission should be almost constant, with the
Comptonized power-law emission varying by as much 
as 60-70\%\footnote{Obviously, it should always be kept in mind that we 
are
here considering only the variability induced by
structural/geometrical changes at fixed total heating rate $L_{\rm h}$. One
should naturally expect, in a realistic situation, that the heating
rate is also variable on a dynamical timescale, and its variations
should contribute to the observed overall variability.}.

The  direct analogy with the light bending model presented above 
lends itself to a simple geometrical
explanation. In the more general framework of two-phase models for the
X-ray
spectra of accreting black holes, the main spectral and variability
properties are determined by the radiative feedback between the hot and
the cold phases. Such feedback is strongly dependent on the geometry
(and on the topology) of the two phases and, in particular, on the sky covering
fraction of the cold phase as seen by the hot, Comptonising medium.
Reflection dominated spectra are expected when the cold phase
intercepts most of the photons coming from the hot phase. This, in the
light bending model, is achieved via general relativistic (GR)
effects, while here it is a result of the clumpy and inhomogeneous nature
of the inner disc.

In more general terms, one can define as ``open'' any geometry in which
the hot, X-ray emitting
plasma is photon starved (i.e. patchy coronae, inner ADAF + outer discs
etc.). Such a geometry will produce hard X-ray spectra, little soft
thermal emission and weak reflection component (see e.g. the simple
model of Zdziarski, Lubi{\'n}ski \& Smith 1999).
On the other hand, a ``closed'' geometry will instead correspond
to a very large covering
fraction of the cold phase, with associated strong soft emission,
softer spectra and strong reflection fraction (Collin-Souffrin et
al. 1996).
Such a geometry
corresponds, observationally, to the highly accreting black
holes. Whether this is due predominantly to strong GR effects as
expected from centrally concentrated coronae atop a cold, thin disc in
the immediate vicinity of the black event horizon, or to the more
mundane inhomogeneous structure of a radiation pressure dominated
disc, remains to be tested with accurate spectral modeling.

Is it possible to discriminate between these two scenarios for a
``closed'' geometry?
In principle, the relativistic blurring induced by the differential rotation of
the inner disc should always be taken into account when fitting
observed spectra. In the original light-bending model, where the
illuminating source is a point-like source above a standard
geometrically thin disc, the ratio of the reflected component to the
power-law continuum is determined by the same effects that determine
 the shape of the relativistic lines, while if the disc is truly
 inhomogeneous, the two effects can be decoupled. Therefore,
 simultaneous spectroscopic studies of relativistically blurred
 emission lines and of the broad band continuum and variability could
 be effectively used to discriminated between a pure light bending
 model and a clumpy disc. Detailed predictions for the latter,
 however, require the combination of sophisticated MHD and radiative
 transfer simulations.

\section{Numerical Simulations}\label{sec:simu}

The toy model presented in Sec.~\ref{sec:toy} provides important qualitative 
predictions regarding the strength and variability of the different
spectral components
emerging from a clumpy accretion flow. In this Section, 
we perform detailed radiative transfer calculations 
to investigate quantitatively the shape of the predicted spectra.
 These calculations account simultaneously for energy balance
 and Comptonisation in the hot phase, together with reflection, reprocessing,
ionization and thermal balance in the cold phase.
The radiative processes in the hot gas are simulated using
the Non-Linear Monte Carlo Code of Malzac \& Jourdain (2000), the reprocessed
radiation from the cold clouds is computed using the photo-ionisation code of
 Ross \& Fabian (1993,2005). The results of each code are used as
 input for the other code and
 self-consistency is achieved by multiple iterations as described in
 Malzac, Dumont \& Mouchet (2005).
 
 \begin{figure}
\psfig{figure=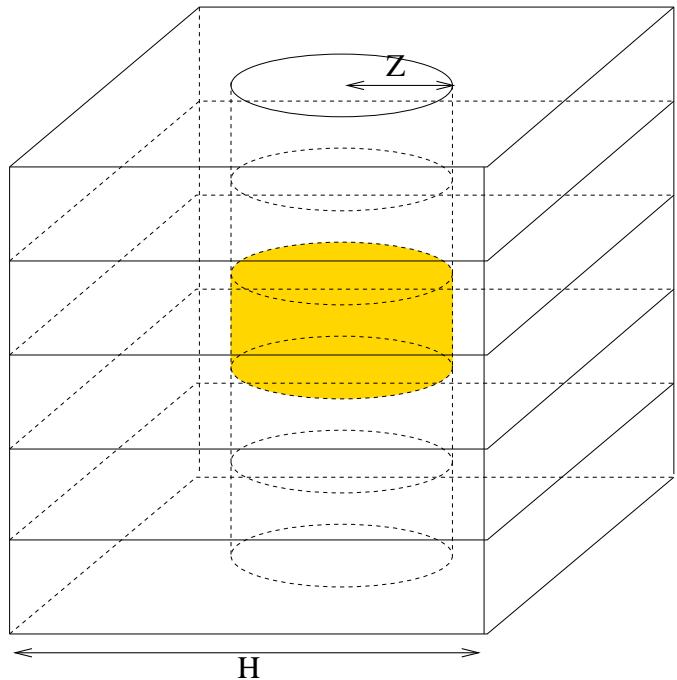,width=0.5\textwidth}
\caption{In numerical simulations, the disc is modeled as an infinite slab 
made of a paving of identical cubic regions. 
As shown in this sketch, each cube is itself divided
into 10 cells to account for inhomogeneity.
The cube is uniformly filled with hot/ionized plasma and cold clumps, 
but dissipation occurs only in the central (colored) zone.}
\label{fig:geom}
\end{figure}

\subsection{Setup}
The geometry we consider for the purpose of our numerical simulations
differs 
from that of the simple analytical model presented in
Section~\ref{sec:toy}.
The basic idea behind our calculations is to study the observed
spectra emerging from a clumpy accretion flow in which the heating is
localized. Thus, the schematic two-zone toy model discussed in Section
2 represents here a patch of the accretion disc, with $H \sim$ disc
scaleheight (spherical symmetry and concentric zones were assumed
there only to allow an exact 
analytic treatment of the two zone model). 
In the simulations, the accretion disc is modeled as an infinite slab made
 of a paving of identical cubes of size $H$ and dissipation takes place
  only close the center the cubes.
 The hot plasma density is uniform and characterized by a 
 Thomson optical depth $\tT=n_{e}\sigma_{\rm T}H$. The disc is also
 uniformly pervaded by cold dense clumps with an effective cloud
 optical depth 
  $\tB$. 
In the case of spherical clumps  the total number of clumps in the cube $N$
   is given by $\tB=\pi N \epsilon^{2}$ (see MC02).
In this idealization, we can afford to simulate the emission from only 
one paving stone by assuming that the same radiation flux 
that escapes to  a neighboring cube  through 
one side, re-enters on the opposite side. 

To account for expected gradients of temperature of the hot plasma and 
ionization parameter of the cold clouds, the cube is divided into ten cells. 
Those cells are defined as the intersection between 5 equidistant
planes that are parallel to the accretion disc and a cylinder the axis
of which is normal to the slab-disc
  and crosses the center of the cube (see Fig.~\ref{fig:geom}).
The plasma is heated only in the central cylindrical cell. The temperature
 in each cell is computed from the local energy balance.
  For a given volume averaged,
 ionisation parameter $\xi$,  the ionisation gradient is computed
  from the local radiative energy density.
  The compactness of the active region is parametrized through
  $h=Z/H$, where 
  $Z$ is the radius of the cylinder.
  We assume that the cloud have
   a constant density, $n_{\rm cl}= 10^{15}$ cm$^{-3}$, 
   and standard (solar) abundances (see Ross and
   Fabian 2005, and references therein for the atomic data used in the
   code). The values for the cold clouds density and ionization
   parameters are therefore chosen within the range of values expected
   for accretion flows around supermassive black holes (see Appendix
   A). The explicit scaling with mass, and the possible relevance of
   our model for stellar mass black holes are discussed in \S~\ref{sec:mass}.
 
 \subsection{Results}
\label{sec:results}
\begin{figure*}
\psfig{figure=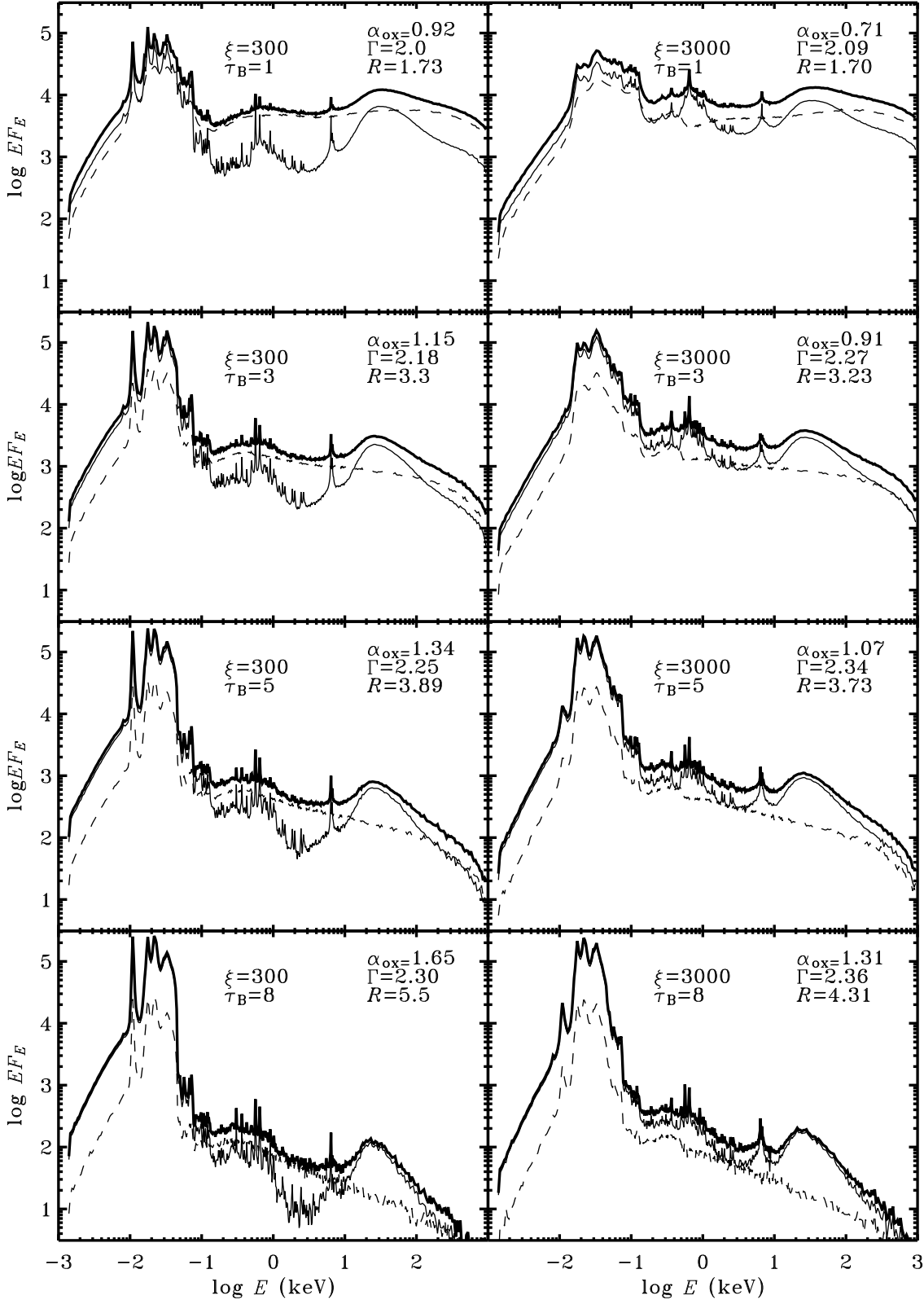,width=0.85\textwidth}
\caption{Effects of $\tB$ and $\xi$ on the emerging spectrum.
Angle averaged spectra calculated for $h=0.1$ and $\tau_{\rm T}=1$.
In each panel, are shown the corresponding $\tB$ and $\xi$ as well as
the best fit parameter $R$ and $\Gamma$ obtained when these spectra
are fitted with
 {\tt PEXRAV} in the 2-30 keV range and the optical to X-ray spectral
 slope $\alpha_{\rm ox}$. 
The thin dashed and solid curves indicate the Comptonised
and   the reprocessed spectra respectively. The total spectra 
are shown by the thick solid curve.}
\label{fig:spectaub}
\end{figure*}

 Fig.~\ref{fig:spectaub} shows a sample of model spectra that
 illustrates the effects of varying $\tB$ from 1 to 8.
The emerging spectra are clearly very complex, with prominent
 signatures of ionized reflection
 dominated by a complex of UV and soft X-ray emission lines.
Strong ionization edges are observed for low
 ionization parameter (for high values of $\xi$ Comptonization has a
 relatively large influence on the emerging spectra and help smearing
 out ionization edges and lines).

\begin{figure}
\psfig{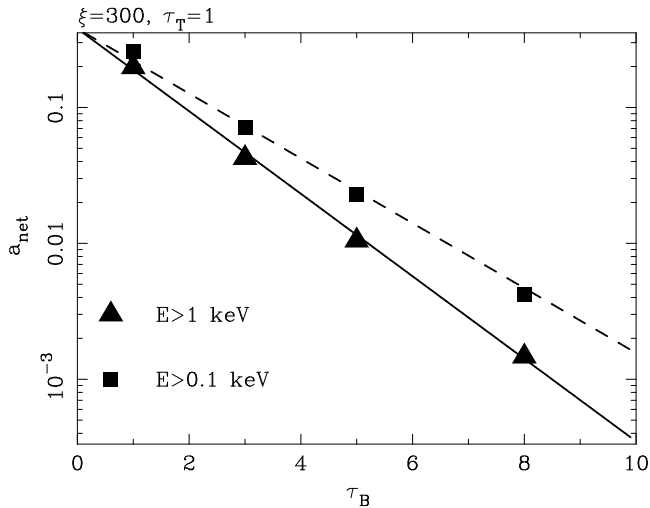}
\caption{Solid symbols show the net albedo of the cold clumps calculated from our simulations, after multiple reflections are taken into
  account. With triangles we show the results
  obtained by considering only the reprocessed radiation above 1 keV,
  while squares show the results obtained by considering reprocessed
  radiation above 0.1 keV. Solid and dashed lines show the
  approximations to the analytical results by the analytic expression
  (\ref{eq:anet}), with $a=0.36$ (equal in both cases), 
  $\ep=\exp(-\tau_{\rm B}/1.43)$ and
  $\ep=\exp(-\tau_{\rm B}/1.82)$, respectively. The solid line also
  shows the relation used in the analytic two-zone toy model
  described in Section~\ref{sec:toy}.}
\label{fig:albedo}
\end{figure}

As mentioned earlier, the complex geometry of our setup will
inevitably lead to multiple reflections, which in turn reduce the net
albedo of the cold clumps. We have studied in detail this effect by
calculating the total albedo as the ratio of incident (Comptonized)
flux to the reprocessed one at energies above 0.1 and 1 keV. The
results are shown in Figure~\ref{fig:albedo}. 
%We found that the effect
%of multiple reflections can be approximated with the 
%expression of eq.~(\ref{eq:anet})
%where $a\simeq 0.36$ and $\ep=\exp(-\tau_{\rm B}/1.43)$ if we consider only the
%reprocessed radiation above 1 keV, and $\ep=\exp(-\tau_{\rm B}/1.82)$ 
%if we consider instead the
%reprocessed radiation above 0.1 keV.

In order to quantify at least the most prominent characteristics of
 these spectra in the X-ray band, we have fitted them
in the 2--30 keV range under {\tt XSPEC} 
with the {\tt PEXRAV} model plus a Gaussian line. 
During the fit the high energy cut-off energy
$E_{c}$ and the inclination angle  were fixed at 400 keV and
30~degrees respectively, and the abundances were fixed to solar.
Obviously, we are not interested here in the goodness of these fits,
 rather we want to use the fit parameters as measures of the spectral
 complexity. The results of the fitting procedure
confirms that reflection dominated spectra (in the sense usually
 implied by an observer, i.e. spectra with $R>1$, where $R$ is the
 value of the reflection fraction parameter of the {\tt PEXRAV} model)
 can be produced with our model for the inhomogeneous disc structure.
For $\tB=8$ the reflection fraction is $R\sim$4--5,
even at large ionisation parameters ($\xi\sim$ 3000).
When comparing these results with the simple toy model of
 Section~\ref{sec:toy}, it is apparent that the {\tt PEXRAV} fits indicate
somewhat larger reflection fractions at small $\tB$.

Below $\sim 1$ keV, all spectra show a distinct curvature that, when
plotted as a ratio to the best fit power-law in the 2-10 keV range,
would appear as a  so-called ``soft excess'', indeed a ubiquitous feature
 in the spectra of QSOs and Seyfert galaxies (see e.g. Wilkes \& Elvis 1987;
 Piconcelli et al. 2005). Within our model, such emission is accounted
 for by a plethora of emission lines produced by ionized reflection of
 the Comptonised continuum off the cold clumps as proposed earlier by
 Ballantyne, Iwasawa and Fabian (2001), Fabian et al. (2002), and more
 recently by Crummy et al. (2005). 

\begin{figure}
  \psfig{figure=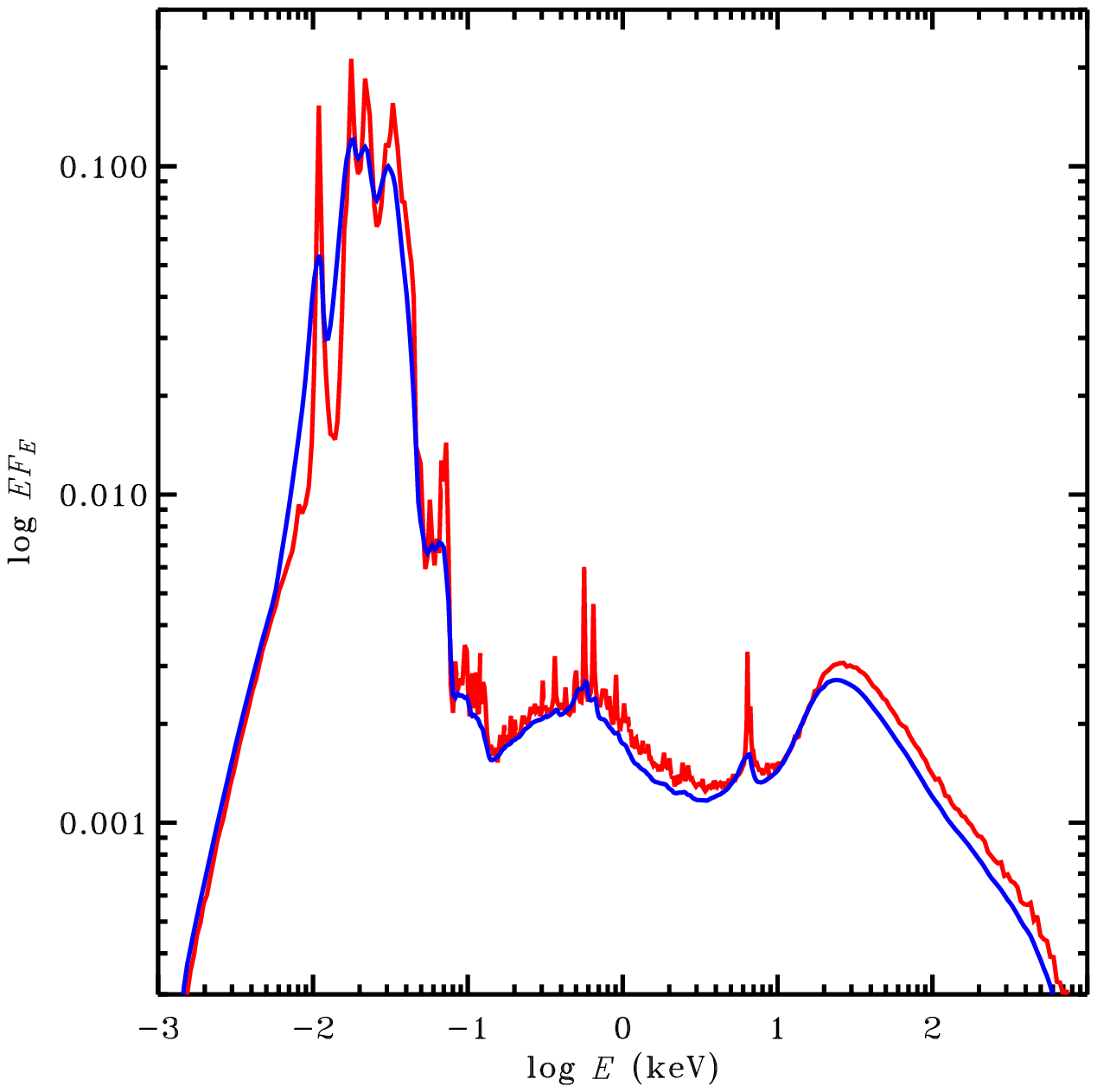,width=0.45\textwidth}
  \psfig{figure=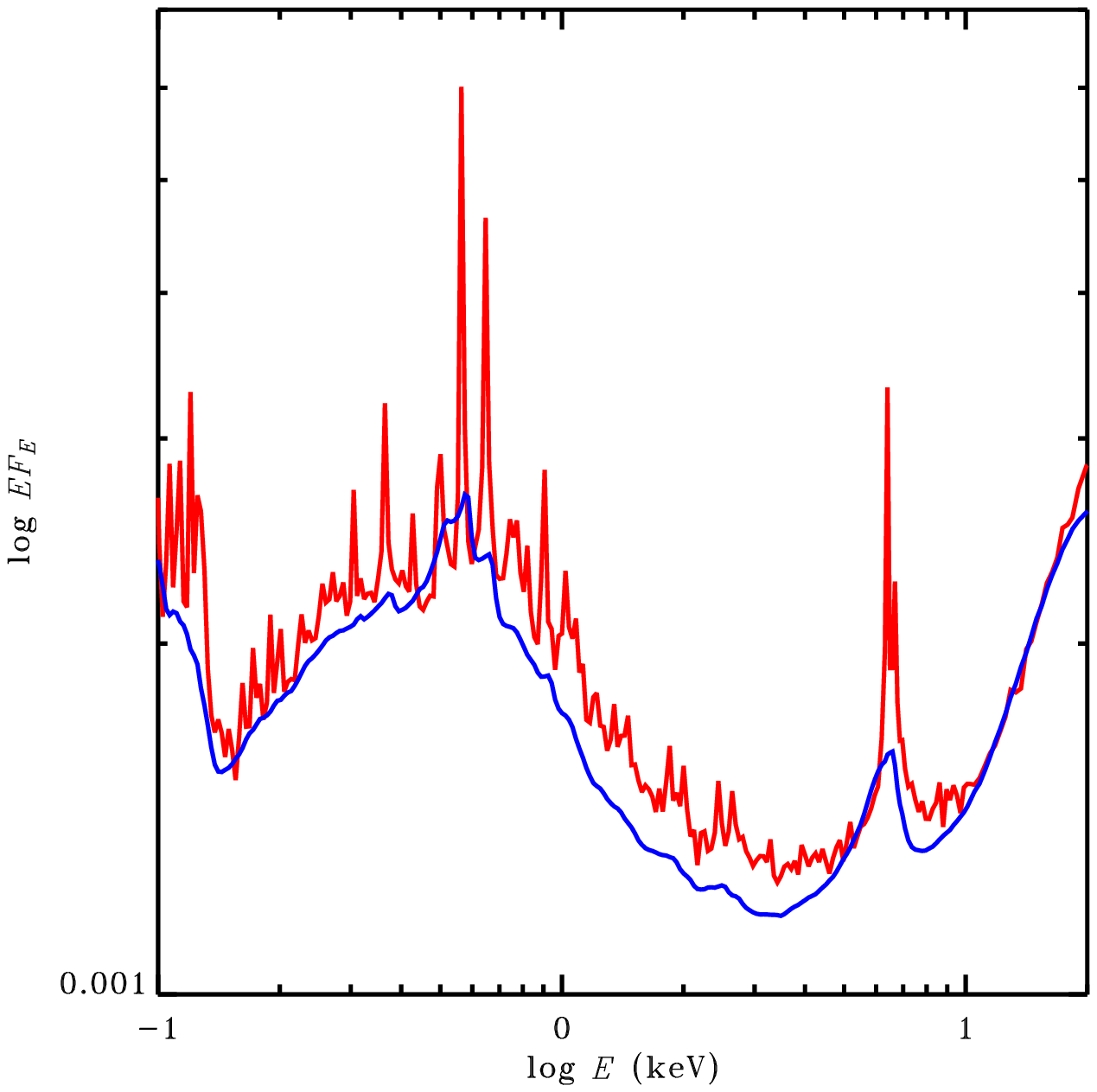,width=0.45\textwidth}
  \caption{The spectrum obtained for $\xi=300$ and $\tB=3$ (red), and
  the same
    spectrum convolved with a relativistic disc model
    (in blue, {\tt kdblur} with $R_{in}=4.5 R_{\rm g}$, $R_{out}=100
    R_{\rm g}$,
    emissivity index 3, inclination of 30 degrees).
    The top panel shows the broad band ratio spectra (1 eV-1 MeV). 
    The bottom panel focuses on the 0.1--30 keV range where most X-ray
    instruments operate.}
\label{fig:kdblur}
\end{figure}

Figure~\ref{fig:kdblur} shows in red a detailed view of the emerging
spectrum for the case $\tB=3$, $\xi=300$. By looking at the ionization
structure within the gas, it is apparent that the gas is still highly
ionized at the outer surface, with the dominant ions being C VII, N
VIII, O IX, Ne X, Mg XI, Si XII, S XIV and Fe XVIII. 
However, the ionization decreases rapidly with depth, 
and by a Thomson depth of 1.4, the least-ionized species 
treated dominates for each element. Specifically 
they are, with their ionization potentials: 
C III (47.9 eV); N III (47.5 eV); O III (54.9 eV); Ne III
(63.5 eV); Mg III (80.1 eV); Si IV (45.1 eV); S IV (47.3 eV) and 
Fe VI (99.1 eV). Within our code, the lowest absorption edge treated 
is thus at 45 eV. Below that energy, 
the only absorption treated is free-free absorption, and radiation can
enter and leave the gas with ease. We should note here that in 
the code helium is assumed to be fully ionized (so that
its only contribution is to the free electron density).
However, since the least-ionized species of C, O and Si
appear to be present in this particular model,
some He II might be present in reality, which would produce
an edge at 54.4 eV if it were included in the calculation. The four 
strongest low-energy emission lines are due to Si IV (11.0 eV), 
O III (17.6 eV), C III (21.6 eV), and O III (33.1 eV). 

The spectra shown in Figure~\ref{fig:spectaub} are those emerging from
a system at rest. It is obvious that, when considering the
application of such a model to the inner region of an accretion disc,
the effects of both the turbulent motion of the clumps and the overall
rotation pattern of the two-phase disc should be taken into
account (as well as cosmological redshift, obviously). 
Indeed, in all previous direct applications of ionized
reflection models to observed data (see e.g. Fabian et al. 2002;
Crummy et al. 2005), the absence of sharp emission lines both in the
soft X-ray and in the Iron K$\alpha$ line regions of the observed
spectra was taken as an indication of extreme relativistic blurring in
the inner region of the accretion disc. Without attempting to address
the issue of including relativistic effects into our model 
in a systematic way, we show here in Figure~\ref{fig:kdblur}
one of our simulated spectra convolved with a relativistic disc
model. The relativistic blurring routine {\tt kdblur} 
(Fabian \& Johnstone, priv. comm.) 
makes use of the {\tt LAOR} kernel describing
relativistic effects on the spectral shape resulting from emission in
an accretion disc orbiting a maximally rotating 
Kerr black hole (Laor 1991). Indeed, even for our
relatively modest choice of blurring parameters (emissivity index $=3$ and
inner radius of the disc at $4.5 R_{\rm g}$), most of the discrete emission
features in the X-ray energy range appear strongly blurred and hardly
distinguishable from a curved continuum. A more extensive analysis of the
strong relativistic corrections to our inhomogeneous model is beyond the scope
of the present paper and will be presented elsewhere. 

In terms of global spectral energy distributions, the general trend
follows the expectations of the simple analytic two-zone model
presented in Section~\ref{sec:toy}. Strongly inhomogeneous discs with
large $\tB$ are dominated by reflected/reprocessed UV emission.
To quantify this, we have calculated the $\alpha_{\rm ox}$ parameter,
i.e. the slope of spectrum between 2500 \AA$\,=5$ eV  and 2 keV: 
$\alpha_{\rm ox}= -0.3838 \log(F_{\rm 2keV}/F_{\rm 2500})$ and the
resulting values are shown in each panel of
figure~\ref{fig:spectaub}. We see a trend of larger $\alpha_{\rm ox}$
for more complex, reflection dominated spectra, as indeed observed
(Gallo et al. 2005).
At face value, $\alpha_{\rm ox}$ 
depends quite sensitively on $\tB$, thus on the degree of
inhomogeneity, and can in principle represent a very useful way to tie down
the general properties of the observed emission. Our model 
would then predict that the
larger $\alpha_{\rm ox}$, i.e. the X-ray weaker a source is, the larger the
reflection fraction should be, and in general, the more complex the X-ray
spectra would appear. It has to be kept in mind, however, that the observed 
emission at 2500 \AA$\,$ may include a contribution from an outer, colder
homogeneous accretion disc. If the extent of the inhomogeneous region of the
accretion disc coincide with the region where radiation pressure dominates,
the relative contribution to the optical emission at 2500\AA$\,$ should
depend on the accretion rate: the larger the accretion rate, the larger is the
contribution from the inner inhomogeneous part we have discussed
here. A full model for the dependence of  $\alpha_{\rm ox}$ on the
source luminosity should then be made based on the combination of a
inner inhomogeneous disc (Appendix~\ref{app:model}) and an outer
standard, gas pressure dominated one.
 
As expected, the reflection features are weaker
 when $h$ or $\tT$ are larger. To confirm this, we have performed two
 additional simulations (not shown in Figure~\ref{fig:spectaub}), and
 again fitted the emerging spectra with {\tt PEXRAV}. The results as shown in
 Table~\ref{tab:fits}. If we double the size of the inner heated region
 (i.e. we take $h=0.2$), while keeping $\tB=1$ constant, the measured
 reflection fraction, for example, drops from 1.73 to slightly less than unity.
 
\begin{table}
      \caption[]{Spectral parameters obtained from the fit of the
 simulated spectra that illustrate the effects of varying the heating
 inhomogeneity parameter $h$ and the Thomson optical depth of the hot
 phase $\tau_{\rm T}$.
In all the simulations $\tB=1$, $\xi=300$. 
$\Gamma_{p}$ and $R$ are the {\tt PEXRAV} photon index and 
reflection amplitude, while EW is the 
equivalent width in eV of the Gaussian line.}
 \begin{tabular*}{0.99\columnwidth}{@{\extracolsep{\fill}}ccccc}
         \hline
            \noalign{\smallskip}
 $h$&  $\tau_{\rm T}$   & $\Gamma_{p}$ & $R$ & EW \\
            \noalign{\smallskip}
            \hline
            \noalign{\smallskip}
             0.1     & 3 &  1.95 & 1.03 & 108 \\
             0.1     & 1 &  2.0   & 1.73  & 82   \\
             0.2     & 1 &  1.98 & 0.98 &   144\\
                      \hline
         \end{tabular*}
\label{tab:fits}
   \end{table}

\section{Discussion}
\label{sec:discus}

\begin{figure}
\hfill
 \psfig{figure=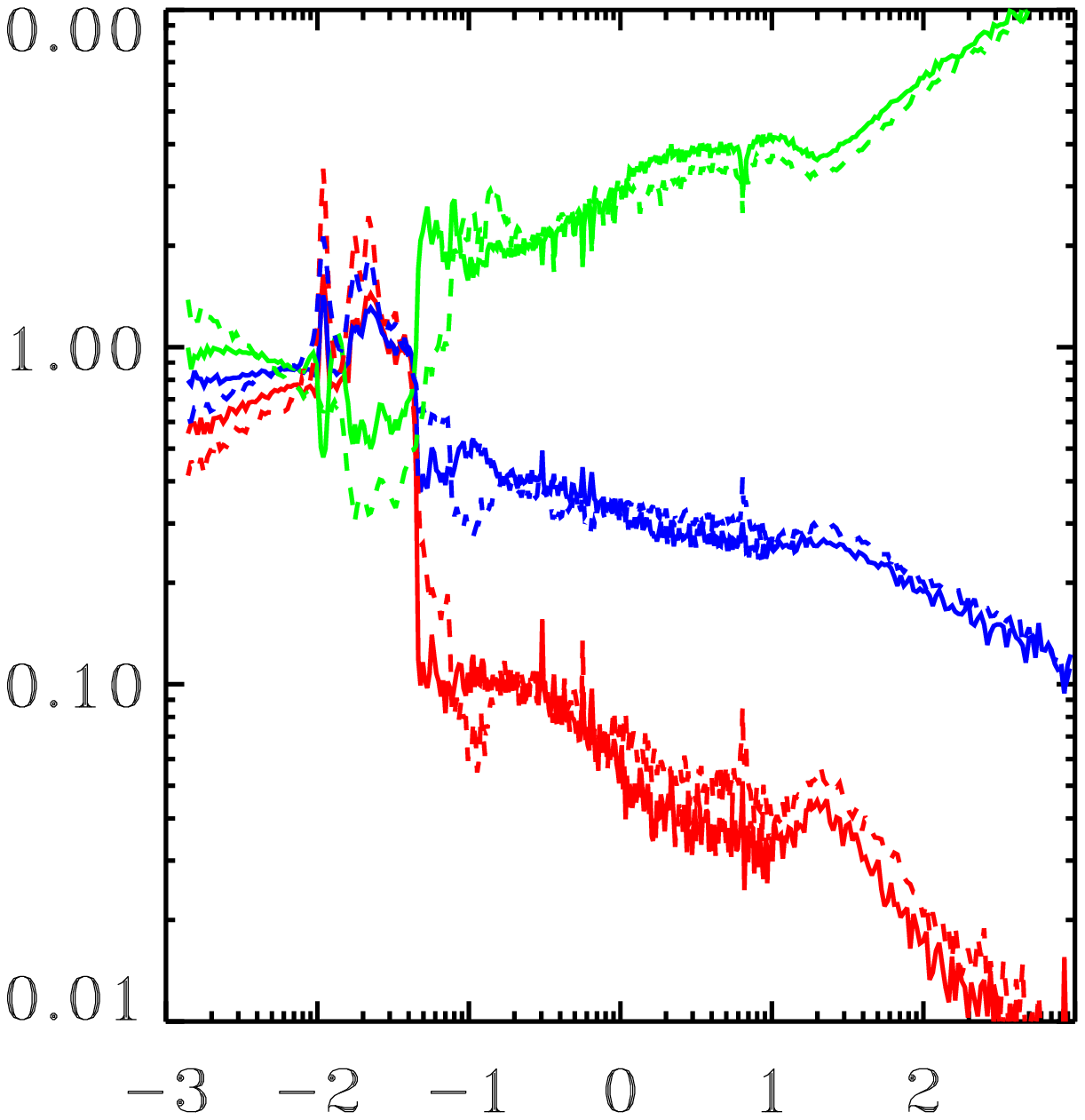,width=0.4\textwidth}
\vspace{0.8cm}

\hfill\psfig{figure=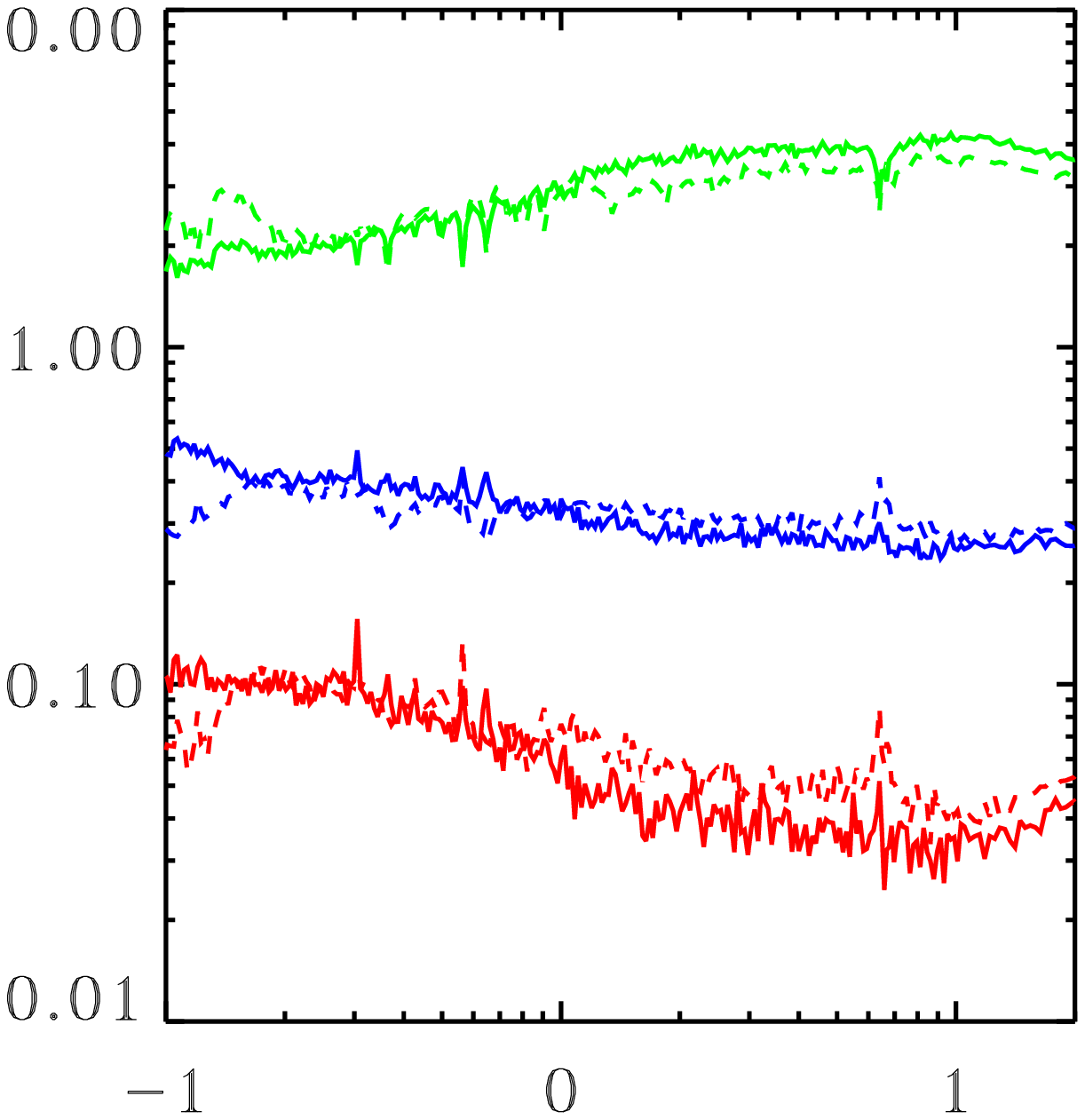,width=0.4\textwidth}
\vspace{0.6cm}
 \caption{Ratio of the spectra obtained for $\tB=1$
 (green), $\tB=5$ (blue) and $\tB=8$ (red) to the spectrum for
 $\tB=3$. The top panel shows the broad band ratio spectra (1 eV-1
 MeV). Solid lines are for $\xi=300$, dashed lines for $\xi=3000$.
The bottom panel focuses on the 0.1--30 keV range where most
X-ray instruments operate. In this limited energy band,
 large changes in $\tB$ can lead to dramatic changes in
luminosity with only weak changes of the spectral shape.}
\label{fig:specvar}
 \end{figure}

The expected variability properties of the inhomogeneous flow were
discussed in Section~\ref{sec:varref}, based on the analysis of our
simple two-zones toy model. Barring any
additional (likely) variability associated with an intermittent
heating rate, we have shown there that a modest 10\% variation in the
optical depth of the cold clouds around a mean value of $\tB\sim$ 1--2 can
produce  large amplitude variations (up to 40--50\%) in the X-ray
power-law flux with much smaller (less than 20--25\%) variations
associated with the reflected component (see figure~\ref{fig:var}). 
This is a possible
explanation for the variability pattern observed in many AGN
 on short (dynamical) timescales, as discussed in more detail in
 Section~\ref{sec:compar}. 
Of course, as in the geometrically
equivalent light bending model of Miniutti \& Fabian (2004), also in our
model there are different regimes of variability, essentially
depending on the degree off inhomogeneity of the heating rate, $h$, and
on the optical depth of the cold clumps $\tB$: when the latter is
large, most of the X-ray emission {\it and} of the reflected radiation
are hidden from view, and large amplitude variability is induced in
both spectral component. 
On longer (thermal, viscous) timescales, the global geometry may
change substantially. Large variations of $\tB$ would of course
correspond to substantial spectral variability. In
figure~\ref{fig:specvar} we have plotted the ratio of the spectra
obtained for $\xi=300$ and $\tB=1,5,8$ to the $\tB=3$ spectrum. Apart
from the dramatic variation of the lowest absorption edge at $\sim$ 45
eV, and of the
overall slope of the Comptonized X-ray emission (getting steeper for large
$\tB$), the spectra
show relatively little variability despite large flux variations,
especially in the X-ray energy range (0.1--30 keV). In particular, no
strong soft excess below 1 keV is seen in the ratio spectrum.

\subsection{Comparison with earlier theoretical works}
Ballantyne, Turner \& Blaes (2004) and Ballantyne, Turner \& Young (2005)
have studied X-ray reflection from inhomogeneous discs. The
vertical density structure of the disc they assume is derived from a
two-dimensional, time-dependent numerical radiation
magneto-hydrodynamical (MHD) calculation. However, their
numerical setup was such that the source of X-ray illuminating
radiation was assumed to be external to the disc (a corona above it),
and dynamically decoupled from it. Here we have tried to address the
different (and computationally much more challenging) situation in
which there is no external hot corona (and no external energy source),
but the heterogeneous, two-phase structure of the flow itself gives
rise to all the observed high energy radiation spectrum.
The spectroscopic studies of Ballantyne and coworkers, however, suggest that
any spectrum produced by a realistic, time varying inhomogeneous accretion
disc, will pose serious interpretative challenges to planned future
reverberation mapping observations. We believe that 
the exploration of ever more
realistic model of inhomogeneous discs, possibly including both the density
structure from 3-D MHD simulations and the inhomogeneous heating and the
detailed radiative transfer calculation similar to those we have presented here
will be an extremely  useful tool for the next generation of X-ray 
spectroscopes.

One of the basic assumptions common to many of the different inhomogeneous solutions
for accretion discs (see e.g. Krolik 1998; Begelman 2002; see also the
discussion in Appendix~\ref{app:model}) is that the clouds
are not confined by magnetic field, but, on the contrary, that
magnetic field lines connect the two phases. This implies that, for
small enough clouds, each of them will be in pressure equilibrium with
the surrounding hot phase. Then, the density contrast (which
is a physical quantity easy to derive from numerical MHD simulations)
will be approximately given by the ratio of the temperatures of the
two phases, which could in principle be inferred from accurate spectroscopic
studies in both soft (0.1-2 keV) and hard (5-30 keV) X-rays. 
In general, a more realistic density structure that takes
also into account pressure equilibrium between the phases in the
presence of a turbulent magnetic field should be  the next step, although
an extremely complicated one, in the implementation of our
inhomogeneous model.

The coupled radiative transfer code we have used in this work is
probably the only one used so far capable of self-consistently
calculating the radiative equilibrium between the two phases present
in the innermost region of an accretion disc with such a detail. 
It is however very time
consuming, and still limited in its wider applicability. The setup we
have discussed here, for example, assumes that all clouds are
optically thick, and neglects transmission of radiation across
them. If in more realistic inhomogeneous discs the cold phase is
distributed among clouds spanning  a range of densities, it is likely
that not only reflection off optically thick clouds, but also
transmission of radiation
should play a role in determining the observed spectra. In
general, in these cases we can
expect, for any given value of $\tB$, reduced cooling and harder
spectra. Moreover, the effects of both total and/or partial absorption
(depending on the filling factor of the cold clumps and on their
geometry) will play an important role. This interesting issue is far
beyond the scope of the present paper, which is intentionally aimed at
exploring a very specific physical configuration. We just note here
that a pure ``absorption model'' to explain the soft excess and the
general X-ray spectral shapes of QSOs and AGN has been recently
proposed by Gierli\'nski \& Done (2004b), and a more general and
detailed study of the role of absorption has been presented in
Chevallier et al. (2005).

Finally, we note that more extreme effect of bulk 
Comptonization which arises in the optically
thick plasma in the vicinity of a black hole (Blandford \& Payne 1981a,b; Payne \& Blandford 1981;
Nied{\'z}wiecki \& Zdziarski 2006 and references therein), has been
ignored. However, as in the inhomogeneous flow we envisage the great
majority of the mass is concentrated in the cold phase (even though
most of the volume is filled by the hot one), this effect should be
safely negligible.

\subsection{Comparison with observations: the case of NLS1}
\label{sec:compar}
Narrow Line Seyfert 1 galaxies, identified by the unusual
narrowness of the broad component of their H$\beta$ lines, are
believed to be powered by SMBH of relatively small masses, with
high accretion rates, possibly close to the Eddington level (see
e.g. Pounds, Done \& Osborne 1995; Boller, Brandt \& Fink 1996).
This fact already makes them very good candidate systems for
inhomogeneous accretion. Furthermore, NLS1 have long been known to be
characterized by  extreme properties of their
X-ray emission: a strong soft excess in the {\it ROSAT} soft band
(0.1-2.4 keV; Boller et al. 1996); unusually steep X-ray spectra
in the hard X-ray band (2-10 keV; Brandt, Mathur \& Elvis 1997; Leighly
1999b, Vaughan et al. 1999); very rapid and large variability
(Leighly 1999a).

Recent XMM detailed spectral studies
have revealed more unusual spectral properties, most notably in the
form of sharp spectral drops above 7 keV, the most extreme examples of
which are found in 1H 0707-495 (Boller et al. 2002) and in IRAS
13224-3809 (Boller et al. 2003). These features are sharp and
time-variable (Gallo et al. 2004).
Detailed spectral modeling of these data so far suggested that either
partial covering (Tanaka et al. 2004; Gallo et al. 2004)
or ionized reflection dominated discs (Fabian et al. 2002),
with light-bending effects (Fabian et al. 2004; 
Crummy et al. 2005)
can explain the observed features in these as well as in other bright
Seyfert galaxies (e.g. 1H 0419-577, Fabian et al. 2005; MCG-02-14-009, 
Porquet, 2005). Within the partial covering model,
variability is induced by rapid changes in the covering fraction of
the absorbers. In the light bending model, instead, variability is
essentially produced by a change in the distance between the X-ray
emitting flares and the reprocessor very close to the black hole.

Our model falls within
the family of models trying to explain the complex X-ray spectra of
bright AGN with ionized reflection, though representing an alternative,
specific, physically motivated, geometrical setup for the multi-phase
structure of the inner regions of an accretion flow near the Eddington 
luminosity. 
In Section~\ref{sec:toy}
we have already discussed the qualitative similarities between the
emerging spectra expected from our inhomogeneous flow and the light
bending models. Here we would like to notice a few points concerning
the partial covering model.

First, a comprehensive analysis of {\it XMM} data of
a number of NLS1 with the partial
covering model (Tanaka, Boller \& Gallo 2005; see also Crummy et al. 2005) 
highlights the fact that
the estimated temperature of the soft X-ray/UV emission component (the
soft
excess) is very similar for objects that span more than four orders of
magnitude in luminosity, as already noticed by Walter \& Fink (1993);
Czerny et al. (2003) and Gierli\'nski and Done (2004b). As pointed out
by Ross \& Fabian (2005), the emission in the 0.2-2 keV band due to
lines and bremsstrahlung in the hot surface layer of the cold medium
is a signature of ionized reflection, and can appear in real data as
blackbody emission, provided that the individual lines and edges are
smeared by a large enough amount, so  that the emerging soft excess will
appear featureless (Crummy et al. 2005; Chevallier et al. 2005).
Secondly, the observed rapid and strong variability
of many NLS1 puts tight constraints on the maximum distance the partial
coverers can possibly be at. In fact, as already discussed in detail
in Gallo et al. (2004), within the partial covering model the absorber
must be outflowing, have a high column density and a low ionization,
which imply a kinetic energy outflow almost two orders of magnitude
larger than the observed luminosity. It is more reasonable to
assume that the absorber is in fact embedded within the X-ray emitter
and shares with it the high rotational velocity of the inner accretion
flow, as we have discussed here.

Naturally, reflection dominated inhomogeneous flows from the vicinity
of a black hole should produce  strongly relativistic
lines in the soft X-ray range. Previous claims of detections of such
lines in the XMM spectra of two Seyfert galaxies (Mkn 766 and
MGC--6-30-15, Branduardi-Raymont et al. 2001; Sako et al. 2003), have
however been questioned (Ballantyne, Ross \& Fabian, 2002; Turner et
al. 2003a) and it is yet unclear to what extent weak relativistic lines
can be seen in spectra clearly dominated by (dusty) warm absorbers
(see Lee et al. 2001; Turner et al. 2001). 

In a recent work, Chevallier et al. (2005) have compared absorption
(and partial covering) and reflection  models with the X-ray spectra
of PG quasars. They reached the conclusion
that absorption models seem capable
of explaining some of the soft X-ray excesses, provided the
absorbing clouds span a range of optical thicknesses, ionization
parameters and covering fractions. On the other hand, ionized
reflection models necessitate that the continuum emission is strongly
suppressed, as in our inhomogeneous model or in the most extreme
versions of the light bending model. 

\subsection{The scaling with the black hole mass}
\label{sec:mass}

The scaling of the physical quantities most relevant for the
observable characteristics of inhomogeneous accretion flows are shown
explicitly in Appendix A. As the solution we describe is thought to
be the final outcome of non-linear MHD instabilities in radiation
pressure dominated discs, it is hard to predict how the parameters
describing the inhomogeneity, $\tau_B$, $\epsilon$ and $h$ will scale
with mass. This is in fact the reason why we have chosen them as free
parameters of the model. However, once their values are fixed (within
the range we have explored here), it is apparent that the hot phase
temperature is almost independent on black hole mass
(eq. ~\ref{eq:theta}),as it was already noticed by Krolik (1998). In
general, the properties of the hot phase turn out to be very
insensitive to the central object mass. For example, the ratio of gas
to radiation pressure in the hot phase does not depend on $m$,
ensuring the analytic treatment presented in appendix A remains valid
throughout the mass spectrum.
As for the other relevant physical quantities, both cold cloud
temperature and ionization parameter scale with mass as 
$m^{-1/4}$, while $n_{\rm cl}\propto m^{-3/4}$: stellar mass
black holes, as expected, have denser, hotter and more highly ionized
clumps.

If one considered also the evaporation/condensation equilibrium
between the phases (neglected here), as done by Krolik (1998), one
would infer that the cold phase covering factor should rise towards
unity as the black hole mass decreases. This correspond to very large
values of $\tau_{\rm B}$, for which the observational differences between an
inhomogeneous flow and a more standard optically thick homogeneous
disc become less and less pronounced.

Finally, there is a more general issue regarding the very existence of
strong inhomogeneities in the accretion flows close to the Eddington
level. In fact, the critical accretion rate above which standard
Shakura-Sunyaev discs are viscously and thermally unstable depend both
on the nature of the viscosity prescription and on the mass of the
central accreting object (see Merloni and Nayakshin 2006). The
observational fact that accretion discs in luminous 
black hole binaries in the so-called high state appear rather stable 
close to the Eddington rate, (Gierli{\'n}ski \& Done  2004a), but 
unstable above it (Done, 
Wardzi{\'n}ski, Gierli{\'n}ski 2004) has been in fact used by Merloni
\& Nayakshin (2006) to argue for a modified viscosity law, based on
the results of an extended local stability analysis. However, if
the nature of such a law is unchanged for supermassive black holes,
the critical accretion rate should scale as $m^{-1/8}$. Thus,
accretion flows in AGN should be unstable at the same (Eddington
scaled) accretion rates at which high state BHXRB are observed to be
stable.

\section{Conclusions}
\label{sec:conc}

We have discussed the expected X-ray spectral and variability properties of
black hole accretion discs at high luminosity, under the hypothesis
that radiation pressure dominated discs are subject to violent
clumping instabilities and, as a result, have a highly inhomogeneous
structure. Within this picture, most of the mass is concentrated in (Thomson)
optically thick clouds that fill only a small fraction of the volume,
which is instead occupied by hot, optically thin Comptonizing plasma. 
Such a geometrical configuration was first proposed by Fabian
et al. (2002) as a mean to explain the complex spectral properties of
the extreme NLS1 1H 0707-495. Since then, many more bright AGN (both
Seyfert and QSO) have been modeled with reflection dominated spectra
(see e.g. Crummy et al. 2005).

Here we have presented a physically motivated model for accretion
discs near the Eddington luminosity (Appendix~\ref{app:model}), 
and studied their radiative output by exploring the
space of parameters that describe the proposed geometry (mean optical
depth of the cold phase and degree of concentration of the heating
source) in a simple two-zone analytical model.
Moreover, we have used a sophisticated radiative transfer code to
simulate the radiative equilibrium between the two phases in a
self-consistent way, so that the mutual radiative feedback is fully
taken into account. 

The main results of our work are the following:
\begin{itemize}
\item{for very inhomogeneous configurations, i.e. those which have both
    a large covering fraction in clouds and very concentrated
    dissipation, the emerging radiation can be dominated by
ionized reflection;}
\item{the observed spectra are very complex, with prominent
 signatures of ionized reflection
 dominated by a plethora of UV and soft X-ray emission lines; if the
 blurring due to relativistic motions in the immediate vicinity of a
 black hole is accounted for, the soft X-ray emission will appear as a
 featureless continuum (a soft excess);}
\item{for a large range of parameters, even when contributing a
large fraction of the hard X-ray spectrum, the reflection component is
less variable than the power-law like emission originating from the
hot Comptonising phase, in agreement with what observed in many Narrow
Line Seyfert 1 galaxies and bright Seyfert 1;}
\item{the reflection fraction becomes strongly variable only at very
    low flux levels of the hard X-ray continuum, 
    corresponding to a configuration in which both
    the primary Comptonized continuum and the high energy reflection
    emission are essentially hidden from the observer's view.}
\end{itemize}

The fact that many of these properties are common to the ``light
bending'' model (Miniutti \& Fabian 2004) is not surprising. 
In fact, for a two-phase system in radiative equilibrium, the main spectral 
properties are determined by the radiative feedback between the hot and
the cold phases. Such feedback is strongly dependent on the geometry
(and on the topology) of the two phases and, in particular, on the sky covering
fraction of the cold phase as seen by the hot, Comptonising medium.
Reflection dominated spectra are expected when the cold phase
intercepts most of the photons coming from the hot phase. 

In more general terms, one can define ``open'' any geometry in which
the hot, X-ray emitting
plasma is photon starved. 
Such a geometry will produce hard X-ray spectra, little soft
thermal emission and weak reflection component.
On the other hand, a ``closed'' geometry will instead correspond
to a very large covering 
fraction of the cold phase, with associated strong soft emission,
softer spectra and strong reflection fraction. 
Such a geometry corresponds, observationally, to the highly accreting
black holes and is obtained via general relativistic (GR) effects
in the light bending model,
 while is a result of the clumpy and inhomogeneous nature of the
inner unstable, radiation pressure dominated part of the disc in the
model we have presented here.

In the end, it seems likely that a complete model for the innermost
region of luminous accretion discs around black holes will involve a
two-phase inhomogeneous and turbulent flow moving at relativistic
speed, so that the emerging spectrum will be determined by ionized
reflection off the cold phase, inverse Compton scattering in the hot
phase and possibly warm absorption from an accompanying outflow, all
distorted by general relativistic effects. The
model we have explored here, thus, represents a step towards our
understanding of such a challenging complexity.

\section*{Acknowledgments}
The authors thank the referee, Andrzej Zdziarski for carefully reading
the manuscript and suggesting notable improvements. AM thanks Luigi
Gallo and Kazushi Iwasawa for useful discussions and the CESR,
Toulouse, where part of this work was carried on, for the
hospitality. JM thanks the Aspen Center for Physics for 
hospitality during the final completion of this work. 
ACF acknowledges the support of the Royal Society.

\appendix
\section{A full inhomogeneous accretion disc model}
\label{app:model}
We derive here a full accretion disc model in which the
inhomogeneous structure is introduced through the parameters
$\tau_B$, $\tau_T$,
$\Theta$, $\epsilon$, $h$, as discussed in the paper. We will follow
closely the approach of Krolik (1998) that studied analytically the
overall equilibrium of a clumpy two-phase accretion disc under the
hypothesis that the cold clumps are magnetically connected to the hot
phase. This configuration turns out to be both thermally and viscously
stable. Its energy flow is more complicated than that of 
standard uniform disc as exemplified by the flow chart of 
Fig.1 of Krolik (1999). The hot phase heat comes either from magnetic
field reconnection and thermalization of cloud random motions (either
due to drag or inelastic collisions), while the cold phase is heated
either through cloud-cloud collisions and through reprocessing of the hot
phase radiation. In order to allow a simplified
treatment of the radiative coupling between the phases, both in
our toy model (Section~\ref{sec:toy}) and in the radiative transfer
simulations (Section~\ref{sec:simu}) we have neglected internal
dissipation in the clouds due to their inelastic collisions, and focussed only on their heating by
absorption of radiation emitted in the hot phase. We will keep this
assumption here, equivalent to setting to zero the term $\alpha_{\rm
  C}$ in Krolik (1998). 

We begin by expressing the angular momentum conservation equation in
terms of the cold clumps velocity dispersion $\sigma_{\rm cl}$. Here
we adopt the customary $\alpha$ viscosity prescription of Shakura \&
Sunyaev (1973), independently of the details on how exactly the
stress is produced (see Krolik 1999 for a discussion). In the
following, 
we will indicate the accretion disc
radius in units of the Schwarzschild radius as $r$, and the
accretion rate in terms of the dimensionless quantity $\dot m =
\epsilon_{\rm rad} \dot M c^2/L_{\rm Edd}$ with the radiative
efficiency $\epsilon_{\rm rad}=1/12$ fixed to its Newtonian value, as
appropriate for the inner boundary condition we have adopted. We have: 
\begin{equation}
\label{eq:angm}
\alpha\sigma_{\rm cl}^2=\frac{\Omega \dot M J(r)}{3 \pi \Sigma},
\end{equation}
where $\Omega$ is the Keplerian angular velocity, 
$\Sigma=2\rho H$ and  the function 
$J(r)=(1-\sqrt{r_{\rm in}/r})$, with $r_{\rm in}=3$, 
describes the Newtonian approximation of the no-torque at
the inner boundary condition for a disc around a Schwarzschild black
hole.
The velocity dispersion of the clouds can be expressed in
terms of the hot phase sound speed, $c_{\rm s,h}$:
\begin{equation}
\sigma_{\rm cl}\equiv {\cal M} c_{\rm s,h} \simeq 3.2 \times 10^8
\Theta_{-1}^{1/2} {\cal M},
\end{equation}
where is the temperature of
the hot phase in units of $m_{\rm e}c^2/k$. As discussed in Krolik
(1998), if the mean separation between the clumps is of the order of
the disc scaleheight, the difference in the potential between neighboring
clumps is then big enough to stir them at ${\cal M}\sim 1$. In fact,
within our simple geometrical model of Section~\ref{sec:toy} and for
small enough clumps, we can
estimate the mean clump separation in units of $H$ as $\sim 0.2
\tau_B^{-1/3} \epsilon_{-1}^{2/3}$.

The angular momentum conservation equation (\ref{eq:angm}) can then be
cast as a constraint on the total (cold plus hot phases) average Thomson 
optical depth of the inhomogeneous disc, $\tau_{\rm tot}=\kappa
\Sigma/2$, with $\kappa$ electron scattering opacity, i.e. the
optical depth the disc would have if it was homogeneous:
\begin{equation}
\label{eq:tau_tot}
\tau_{\rm tot}=2.6 \times 10^{3} \alpha^{-1} r^{-3/2} \dot m J(r) {\cal
  M}^{-2} \Theta_{-1}^{-1}\,.
\end{equation}

Let us now discuss the hydrostatic equilibrium of the disc. For the
hot phase, assuming that gas pressure dominates over radiation
pressure, the disc scaleheight is given by:
\begin{equation}
\label{eq:hot}
H\simeq 1.4 \times 10^{10} m_7 r^{3/2} \Theta_{-1}^{1/2}\,.
\end{equation}
The cold clumps, obviously, will be distributed over a scaleheight of
the order ${\cal M} H$.

We can now verify that indeed our assumption about the pressure
dominance in the hot phase is justified. The gas pressure is given by:
\begin{equation}
P_{\rm gas,h}=2\rho_{\rm h} k T_{\rm hot}\simeq 1.9 \times 10^7
m_7^{-1} r^{-3/2} \Theta_{-1}^{1/2} \tau_T\;{\rm dyne}\,,
\end{equation} 
while for the radiation pressure we have
\begin{equation}
P_{\rm rad,h}=\tau_T Q/(3c) \simeq 1.8 \times 10^8 m_7^{-1} r^{-3} \dot
m_{-1} J(r) \tau_T\;{\rm dyne}\,,
\end{equation} 
where $Q$ is the disk dissipation rate per unit
area. Their ratio is then given by:
\begin{equation}
\label{eq:pratio}
\frac{P_{\rm rad,h}}{P_{\rm gas,h}}\simeq 5.8 r^{-3/2} \dot m_{-1}
J(r) \Theta_{-1}^{-1/2} < 0.11 \dot m_{-1} \Theta_{-1}^{-1/2}\,,
\end{equation}
where the inequality in the above expression comes from taking the maximum
of the radial function $r^{-3/2}J(r)$ (at $r=16/3\simeq5.33$). So our
assumption is justified for sub-Eddington disks.

The radiative properties of the system will be in general a function
of the density, temperature, total number and geometry of the cold
clouds. We can estimate the cold clump temperature under the zeroth
order assumption that they will re-radiate thermally all the radiation
coming from the hot phase.
 Consequently, we can write $Q/(ch^2)=4\sigma_{\rm B}T_{\rm bb}/c$, 
where $\sigma_B$ is the Stefan-Boltzmann constant, and 
we have allowed for inhomogeneous heating by introducing the parameter
$h$, the size of the heated region in units of the disc scaleheight
(see Section~\ref{sec:toy}).
Thus we get, for the cold cloud temperature:
\begin{eqnarray}
\label{eq:tbb}
kT_{\rm bb} &\simeq& 1.1 \times 10^2 m_7^{-1/4} r^{-3/4}
[(1-a)/0.64]^{1/4} 
\nonumber \\&\times& [\dot
m_{-1} J(r)]^{1/4} h_{-1}^{-1/2}\;{\rm eV}\,.
\end{eqnarray}
Then, assuming pressure balance between the clouds and the hot ambient
phase we get, for the cloud density:
\begin{eqnarray}
\label{eq:ncl} 
n_{\rm cl}&\simeq& 1.0 \times 10^{17} m_7^{-3/4} r^{-3/4} [(1-a) \dot
m_{-1} J(r)]^{-1/4} \nonumber \\
&\times& h_{-1}^{1/2} \tau_T \Theta_{-1}^{1/2}\;{\rm
  cm}^{-3}\,.
\end{eqnarray}
Neglecting the weak dependence on the albedo $a$, 
the typical ionization parameter, in units of 
  erg cm s$^{-1}$, is given by
\begin{equation}
\xi\simeq  1.3 \times 10^{5} m_7^{-1/4} r^{-9/4} [\dot
m_{-1} J(r)]^{5/4} h_{-1}^{-5/2} \tau_T^{-1} \Theta_{-1}^{-1/2}\;.
\end{equation}
For the fiducial values of the parameters we are discussing here,
$\xi$ attains a maximum value of about $10^{3}$, close to the radius of
maximal dissipation in the disc. Also as expected, the strongest
dependence is on the geometrical factor $h$, as the ionization state
of the cloud surfaces will be most sensitive to the heating inhomogeneity.

The cold clouds filling factor is given by $f=\frac{4}{3} \epsilon
\tau_B$, and the total density in the disc is then
\begin{equation}
n_{\rm tot}=f n_{\rm cl}+(1-f)n_{\rm h}=n_{\rm cl}[f+(1-f)\frac{T_{\rm
    bb}}{T_{\rm hot}}]\approx \frac{4}{3} \epsilon \tau_B n_{\rm cl}\,.
\end{equation}
It is now possible to close the disc structure equations and find a
relationship between the model free parameters ($\tau_B$, $\tau_T$,
$\Theta$, $\epsilon$, $h$) and the physical parameters of the standard
accretion disc solutions (mass, accretion rate, viscosity, etc.). In
order to do so, we equate the total disc optical depth, dominated by
the cold phase, $\sigma_T n_{\rm tot} H$ to the expression in
equation~(\ref{eq:tau_tot}). We obtain, finally:
\begin{equation}
\tau_B \tau_T \Theta_{-1}^2 \epsilon_{-1} h_{-1}^{1/2} \simeq 22
r^{-9/4} [\dot m_{-1} J(r)]^{5/4} m_7^{-1/4} \alpha^{-1} {\cal M}^{-2}
\end{equation}

So far, we have neglected the condensation/evaporation
equilibrium between the cold and the hot phase. If we introduce the
relationship between the typical cloud size and the hot phase
temperature and density, as derived by Krolik (1998, eq.5), we obtain
the following expression for the hot phase temperature:
\begin{eqnarray}
\Theta &\simeq& 0.36 \alpha^{-2/7} m_7^{-1/14} r^{-9/14} [\dot m_{-1}
J(r)]^{5/14} \nonumber \\ &\times& 
{\cal M}^{-4/7} \tau_B^{-2/7} h_{-1}^{1/7}\,,
\label{eq:theta}
\end{eqnarray} 
almost independent on black hole mass and only weakly dependent on
accretion rate.

To conclude, it is interesting to notice that we have obtained 
a self-consistent solution for the inner
radii (i.e. those where radiation pressure is supposed to dominate) 
of supermassive black holes accreting at a few tenths of the
Eddington rate for our fiducial parameters, i.e. both $\tau_T$ and
$\tau_B$ of the order unity and $h$, $\epsilon$, $\Theta$ all of the
order $10^{-1}$. In both Sections~\ref{sec:toy} and~\ref{sec:simu} we
have explored in detail the radiative equilibrium and the expected
spectral and variability signature of an accretion disc in which all
the above parameters are either fixed at, or allowed to vary around,
these fiducial numbers. The fact that they indeed correspond to the
right values for a full accretion disc model as we have outlined here
is indeed a very encouraging result.

\section{Analytical estimates of the emerging luminosities for the
  two-zone model}
\label{app:toy}
For each zone, the escaping luminosity is related to the internal
energy through the escape probability (Lightman \& Zdziarski 1987):
\begin{equation}
\dot P=(L/U)(H/c),
\label{eq:pesc}
\end{equation}
which depends on the geometry, sources distribution, optical depth and
energy.  We consistently indicate as $\pescc^j$,
$\pescs^j$,$\pescr^j$,  the
escape probability for Comptonized, soft and reflected radiation, with
$j={\rm i,o}$ for the inner and outer zone, respectively, and
will use the analytical approximation to $\dot P$ given in the
Appendix of MC02.
For the inner zone we assume for the escape probability the analytic
formula appropriate for a spherical geometry with central isotropic
injection (eq.~A3 of MC02). For the outer zone, a central isotropic
injection in spherical geometry is appropriate as long as $h\ll 1$,
while for $h\longrightarrow 1$ a slab geometry should be more
appropriate. Thus, we employ an interpolating expression for the
intermediate cases.

Also, we should
include a term that depends on the {\it average} fraction $\bar K$
of the emerging luminosity from the
outer zone that reenters the inner one. This is computed as follows:
consider a shell of width $dr$ located at a distance $r$ from the
center, with $r>h$. The solid angle (divided by $2\pi$)
subtended by the inner zone as
seen from the shell is $K(r)\equiv \Delta \Omega/2\pi
=1-\sqrt{1-(h/r)^2}$. Then we can define
\begin{equation}
\bar K = \frac{1}{1-h}\int_h^1 K(r) dr = 1-{\cal I}/(1-h),
\end{equation}
where the integral is evaluated to
\begin{equation}
{\cal I}=\sqrt{1-h^2} + h \arcsin{h} - h (\pi/2).
\end{equation}

As we assume uniform density for the hot and cold phase, we have that
the optical depths in the two zones are simply given by: $\tT^{\rm
   i}=h\tT$; $\tT^{\rm o}=(1-h)\tT$, and analogously for $\tB$.

\begin{enumerate}
\item[{\it Inner Zone:}]
The soft luminosity in the hot phase is produced by the clouds through
absorption and reprocessing of the Comptonized and reflected radiation
and disappears through escape or Comptonization, additionally, a
fraction of the soft luminosity emitted by the outer zone reenters the
inner zone. Thus, the radiative
equilibrium balance for $\Uuv^{\rm i}$ reads:
\begin{eqnarray}
(\pescs^{\rm i}+\tT^{\rm i})\Uuv^{\rm i} &=& \tB^{\rm i}(1-a)\Uc^{\rm
   i}+\tB^{\rm i}(1-\aR )\UR^{\rm i} \nonumber \\ 
&+&(\bar K)\pescs^{\rm o}\Uuv^{\rm o},
\label{eq:usbal_in}
\end{eqnarray}
where $a$ and $\aR$ are  the clouds energy and angle integrated albedo
for a Comptonized and a reflection spectrum, respectively.

Similarly, the reflected radiation is formed through reflection of the
Comptonized and reflected radiation on the clouds, and disappears via
escape, Comptonization in the hot plasma and absorption by the clouds:
\begin{equation}
[\pescr^{\rm i}+\tT^{\rm i}+\tB^{\rm i}(1-\aR)]\UR^{\rm i}=
\tB^{\rm i} a \Uc^{\rm i} + \bar K \pescr^{\rm o}\UR^{\rm o}.
\label{eq:urbal_in}
\end{equation}

The fraction of the internal energy in the form of Comptonized
radiation has as main source the power dissipated in the hot plasma
but also the Comptonized soft and reflected radiation; the sink term
is due to escape, absorption and reflection by the clouds:
\begin{equation}
(\pescc^{\rm i}+\tB^{\rm i})\Uc^{\rm i}= \Lh H/c+\tT^{\rm i}
\Uuv^{\rm i}+\tT^{\rm i} \UR^{\rm i} + \bar K \pescc^{\rm o}\Uc^{\rm
o}.
\label{eq:uxbal_in}
\end{equation}

\item[{\it Outer Zone:}]
Similarly, we can write the equations for the radiative equilibrium in
the outer zone, considering that all the luminosities emerging from
the inner zone have to enter the outer zone:
\begin{eqnarray}
(\pescs^{\rm o}+\tT^{\rm o})\Uuv^{\rm o} &=& \tB^{\rm o}(1-a)\Uc^{\rm
   o}+\tB^{\rm o}(1-\aR )\UR^{\rm o} \nonumber \\ 
&+& \pescs^{\rm i}\Uuv^{\rm i},
\label{eq:usbal_out}
\end{eqnarray}
\begin{equation}
[\pescr^{\rm o}+\tT^{\rm o}+\tB^{\rm o}(1-\aR)]\UR^{\rm o}=
\tB^{\rm o} a \Uc^{\rm o}+\pescr^{\rm i}\UR^{\rm i},
\label{eq:urbal_out}
\end{equation}
\begin{equation}
(\pescc^{\rm o}+\tB^{\rm o})\Uc^{\rm o}= \tT^{\rm o}
\Uuv^{\rm o}+\tT^{\rm o} \UR^{\rm o} + \pescc^{\rm i}\Uc^{\rm i}.
\label{eq:uxbal_out}
\end{equation}
\end{enumerate}

The above system of six equation in the six unknown ($U_i^j$, with
$j={\rm i,o}$ and $i={\rm C,R,UV}$) can be solved analytically, and
the three escaping luminosities finally computed through
eq.~(\ref{eq:pesc}).

An estimate for the observed reflection coefficient can be
directly obtained from:
\begin{equation}
R\sim \frac{\LR^{\rm o}}{a_{\rm net}\Lx^{\rm o}},
\label{eq:R}
\end{equation}
where $\Lx$ is the Comptonized luminosity above 1 keV. 
For the net albedo (accounting for multiple reflections), $a_{\rm net}$, we use the expression in eq.(\ref{eq:anet}), derived from a calibration 
with the numerical simulations, as discussed in Section~\ref{sec:toy}.
This formula corresponds to an angle averaged reference
slab-reflection spectrum.  Actually, the reflection coefficient also
depends on the assumed inclination angle for the infinite slab model
(see e.g. {\sc PEXRAV} model in {\sc XSPEC}, Magdziarz \& Zdziarski
1995).  Equation~(\ref{eq:R}) could be corrected for this, for example
by dividing it by the angular factor given by equation 2 of
Ghisellini, Haardt \& Matt (1994).  However, for the relatively low
inclination angles usually assumed in spectral fits, the correction is
small and, for simplicity, we will neglect it.

%Excluding the reflection component, the total luminosity sinking out
%of the hot phase either through escape or interaction with the clouds
%is, for the inner and outer zones, respectively:
%\begin{equation}
%L^{\rm i}=(\pescx^{\rm i}+\tB^{\rm i})\Ux^{i} 
%+(\pescs^{\rm i}+\tT^{\rm i}+\tB^{\rm i})\Uuv^{i},\label{eq:Li}
%\end{equation}
%and
%\begin{equation}
%L^{\rm o}=(\pescx^{\rm o}+\tB^{\rm o})\Ux^{o} 
%+(\pescs^{\rm o}+\tT^{\rm o}+\tB^{\rm o})\Uuv^{o};\label{eq:Lo}
%\end{equation}
%while the soft luminosities entering the hot phase (coming from the 
%clouds) are given by:
%\begin{equation}
%\Ls^{\rm i}=(\pescs^{i}+\tT^{i}+\tB^{i})\Uuv^{i}+\bar K \pescs^{o} \Uuv^{o} \label{eq:Lsi}
%\end{equation}
%and
%\begin{equation}
%\Ls^{\rm o}=(\pescs^{o}+\tT^{o}+\tB^{o})\Uuv^{o}+\pescs^{i} \Uuv^{i}. 
%\label{eq:Lso}
%\end{equation}
%We then define an average amplification factor $A = (L^{\rm i}+L^{\rm
%  o})/(\Ls^{\rm i}+\Ls^{\rm o})$. The amplification factor is 
%so defined should be related to an average
%photon index $\Gamma$ of the Comptonized photons in the two zones,
% which can therefore be expressed -- using the
%phenomenological formula given by Beloborodov (1999) -- as
%\begin{equation}
%\Gamma=2.33(A-1)^{-1/\delta},
%\label{eq:belo}
%\end{equation}
%where the parameter $\delta=1/10$ for AGN and $1/6$ for GBHs.

\bsp
\label{lastpage}

\end{document}